\documentclass[10pt,aps,prd,onecolumn,showpacs,amsmath,amssymb,nofootinbib,eqsecnum,preprintnumbers,superscriptaddress]{revtex4-2}

\usepackage{xcolor}
\usepackage{graphicx}
\usepackage{comment}
\usepackage{tikz}
\usepackage[colorlinks=true,
            citecolor=red,
            linkcolor=blue,
            urlcolor=violet,
            filecolor=cyan,
            backref=false]{hyperref}
\usepackage{multirow}

\newcommand\feq{\mathrel{\phantom{=}}}

\begin{document}

%%%%%%%%%%%%%%%%%%%%%%%%%%%%%%%%%%%%%%%%%%%%%%%%%%%%%%%%%%%%%%%%%%%%%%%%%%%%%%%%%%%%%%
%% TITLE

\title{
Carrollian limit of quadratic gravity
}

\author{Poula Tadros}
\email{poulatadros9@gmail.com}
\affiliation{Institute of Theoretical Physics, Faculty of Mathematics and Physics, Charles University,
V Hole\v{s}ovi\v{c}k\'ach 2, Prague 180 00, Czech Republic}
\affiliation{Department of Applied Physics, Aalto University School of Science, FI-00076 Aalto, Finland}

\author{Ivan Kol\'a\v{r}}
\email{ivan.kolar@matfyz.cuni.cz}
\affiliation{Institute of Theoretical Physics, Faculty of Mathematics and Physics, Charles University,
V Hole\v{s}ovi\v{c}k\'ach 2, Prague 180 00, Czech Republic}

\date{\today}

\begin{abstract}
We study the Carrollian limit of the (general) quadratic gravity in four dimensions. We find that in order for the Carrollian theory to be a modification of the Carrollian limit of general relativity, the parameters in the action must depend on the speed of light in a specific way. By focusing on the leading and the next-to-leading orders in the Carrollian expansion, we show that there are four such nonequivalent Carrollian theories. Imposing conditions to remove tachyons (from the linearized theory), we end up with a classification of Carrollian theories according to the leading-order and next-to-leading-order actions. All modify the Carrollian limit of general relativity with quartic terms of the extrinsic curvature. To the leading order, we show that two theories are equivalent to general relativity, one to $R+R^2$ theory, and one to the general quadratic gravity. To the next-to-leading order, two are equivalent to $R+R^2$ while the other two to the general quadratic gravity. We study the two theories that are equivalent to $R+R^2$ to the leading order and write their magnetic limit actions.

\end{abstract}

\maketitle

%%%%%%%%%%%%%%%%%%%%%%%%%%%%%%%%%%%%%%%%%%%%%%%%%%%%%%%%%%%%%%%%%%%%%%%%%%%%%%%%%%%%%%
%% INTRODUCTION

\section{Introduction}

The \textit{quadratic gravity} can be derived as an effective field theory by truncating the expansion of the bosonic section of string theory with the first order being \textit{general relativity (GR)} \cite{ZUMINO1986109,ZWIEBACH1985315,Forger:1996vj,Myers:1987yn,Alvarez-Gaume:2015rwa} or by imposing a maximal momentum to strings \cite{Nenmeli:2021orl}. It has been studied even before the connection to string theory as a renormalizable theory of gravity \cite{Stelle:1977ry,Stelle:1976gc,Julve:1978xn}. It admits a wide class of black-hole and other spherically symmetric (exact) solutions \cite{Lu:2015cqa,Lu:2015psa,Podolsky:2019gro,Pravdova:2023nbo}. Nevertheless, in general, it suffers from the presence of unphysical ghost and tachyonic degrees of freedom \cite{Stelle:1976gc}.

The \textit{Carrollian limit} was first considered independently by Levy-Leblond \cite{Levy-Leblond} and Sen Gupta \cite{Gupta} as the ultralocal limit of the Poincar\'e group where the speed of light $c$ approaches zero, ${c\to0}$. However, at the time, due to the lack of physical application of this limit, it was only studied by mathematicians until 40 years later when the Carrollian limit was linked to many applications in physics. Now, Carrollian physics and Carrollian structures are studied in the context of representations of the Carroll group i.e. Carroll particles \cite{Zhang:2023jbi,Bergshoeff_2014,Marsot:2021tvq,Marsot:2022imf}, condensed matter physics \cite{Bagchi:2022eui,Kubakaddi_2021,Kononov_2021}, field theory \cite{PhysRevD.106.085004,Chen:2023pqf,Bergshoeff:2022eog,Henneaux:2021yzg}, conformal field theory \cite{Bagchi:2019xfx,Bagchi:2019clu,Bagchi:2021gai,PhysRevD.103.105001}, fluid mechanics \cite{Bagchi:2023ysc,Ciambelli:2018wre,Ciambelli:2018xat,campoleoni2019two,Ciambelli:2020eba,10.21468/SciPostPhys.9.2.018}, cosmology \cite{deBoer:2021jej,Bonga:2020fhx}, string theory \cite{PhysRevLett.123.111601,Bagchi:2021ban,Cardona:2016ytk}, gravity \cite{Perez:2021abf,Perez:2022jpr,Hartong:2015xda,Figueroa-OFarrill:2021sxz,Hansen:2021fxi,Gomis:2020wxp,Bergshoeff:2022qkx,Guerrieri:2021cdz,Hansen:2020wqw} (it is regarded as the strong coupling limit of gravity theories \cite{Anderson:2002zn}), black holes \cite{Donnay:2019jiz,Grumiller:2019tyl,Redondo-Yuste:2022czg,Marsot:2022imf,Anabalon:2021wjy,Ecker:2023uwm}, null boundaries \cite{Herfray:2021qmp,Chandrasekaran:2021hxc,Bagchi:2019clu,Ciambelli:2019lap} and dynamics of particles near black-hole horizons \cite{Gray:2022svz,Bicak:2023vxs,Marsot:2022qkx}.

The connection between the Carrollian limit and physics near black-hole horizons was shown in \cite{Donnay:2019jiz} utilizing the membrane paradigm \cite{Price:1986yy,1986bhmp.book.....T,Damour:1978cg} which is a paradigm showing that the physics of a black hole on a stretched horizon is dual to that of a relativistic fluid on a ${(2+1)}$-dimensional submanifold. Taking the Carrollian limit of both sides gives a duality between physics on the horizon and a Carrollian fluid. It was shown afterwards that there are two nonequivalent Carrollian limits of a relativistic theory called the \textit{electric} and \textit{magnetic limits}. The electric limit comes directly from the \textit{leading order (LO)} in the \textit{Carrollian expansion}, i.e., the expansion in $c$, while the magnetic limit is a certain truncation of the \textit{next-to-leading order (NLO)} of this expansion.

In this paper we analyze the electric and the magnetic Carrollian limit of quadratic gravity, which is the first step towards the analysis of dynamics of particles near black-hole horizons. We study the electric limit of the general quadratic gravity theory, construct a classification of Carrollian theories from it, and the magnetic limit of the resulting ghost-free theories. Throughout the paper we use the units where Newton's constant $G$ is set to ${G={1}/{(16\pi)}}$. The paper is organized as follows:
\begin{itemize}   
    \item In Sec.~\ref{sec:2}, we review the pre-ultralocal (PUL) parametrization, which is suitable for the Carrollian expansion, and calculate the PUL versions of various tensors appearing in a general four-dimensional quadratic gravity action.
    \item In Sec.~\ref{sec:4}, we review the electric Carrollian limit of GR and show the ultralocality of the spacetime evolution.
    \item In Sec.~\ref{sec:5}, we perform the Carrollian expansion of quadratic gravity action. We show that the parameters $\alpha$ and $\beta$ in the action must depend on $c$ in a specific way otherwise the resulting theory would be drastically different from the Carrollian limit of GR. Requiring the resulting theory to be a modification to the Carrollian limit of GR to LO or NLO gives four nonequivalent Carrollian theories.
    \item In Sec.~\ref{sec:6}, we study those limits one by one and derive conditions on $\alpha$ and $\beta$ to remove tachyons (from the linearized theory) in each case to the LO and NLO.
    \item In Sec. \ref{magnetic limit section}, we study the magnetic limit of the ghost-free and tachyon-free theories.
    \item The paper is concluded with a brief summary and discussions of our results in Sec.~\ref{sec:concl}.
    %\item In Appendix~\ref{sec:3}, we review the mathematical aspects of Carrollian physics from algebraic and geometric points of view and explain the duality to physics near black holes' horizons in more detail.
    %\item In Appendix~\ref{appendix 3}, we review the basics of Carrollian transformations on flat spaceimes, and how they induce symmetries on a general (curved) Carrollian manifold. We also summarize the importance of truncation for the NLO Lagrangians to get Carrollian theories.
\end{itemize}

\section{Pre-ultralocal parametrization}\label{sec:2}

The \textit{pre-ultralocal (PUL) parametrization} is a parametrization of the metric on a manifold using the decomposition of its tangent bundle into a vertical and horizontal subbundles (see below). It is the most convenient parametrization of the spacetime for the analysis of Carrollian gravity since it is well adapted to the ultralocal structure of the Carrollian limit and it displays the speed of light $c$ explicitly, which makes the calculations more obvious. In what follows, we briefly explain the mathematical background of the PUL parametrization. By following the calculations and notations from \cite{Hansen:2021fxi}, we present the PUL version of the Riemannian tensor which will be used to calculate terms in quadratic gravity action in later sections.

Let $(M,\boldsymbol{g})$ be a ${(d+1)}$-dimensional Lorentzian manifold (with mostly positive signature). Let us denote the tangent bundle of $M$ by $TM$ and define two sub-bundles of $TM$ according to the signature of the metric: The first is called the \textit{vertical bundle} $\mathrm{Ver}M$ (or the timelike bundle) and it corresponds to the timelike direction, i.e., its fibers are endowed with a vector space isomorphic to the time coordinate. The second is referred to as the \textit{horizontal bundle} $\mathrm{Hor}M$ (or the spatial bundle) and it represents the remaining $d$ spacelike directions. It is easy to prove that ${TM= \mathrm{Ver}M \oplus \mathrm{Hor}M}$. Furthermore, it generates a foliation of the manifold whose slices are the sub-manifolds of a constant time coordinate $t$. This foliation allows us to define orthogonal spatial and timelike sections as follows: Consider a covector $T_{\mu}$ and a vector $V^{\mu}$ from $\mathrm{Ver}M$, where ${\mu,\nu, \ldots= 1,2, \dots,d+1}$ are tensor indices in $TM$. Next, we consider a symmetric tensor $\Pi_{\mu \nu}$ from $\mathrm{Hor}M$, which is the induced metric (or the first fundamental form), and its inverse $\Pi^{\mu \nu}$.

By construction of the sub-bundles and the foliation we have 
\begin{equation} \label{eq:1}
\begin{aligned}
T_{\mu}V^{\mu} &=-1, 
&
-V^{\mu}T_{\nu} + \Pi^{\rho \mu} \Pi_{\rho \nu} &= \delta_{\mu}^{\nu},
&
T_\mu \Pi^{\mu \nu} = \Pi_{\mu \nu} V^\nu&=0, 
\end{aligned}
\end{equation}
The PUL parametrization of the metric $g_{\mu\nu}$ is given by 
\begin{equation} \label{eq:3}
\begin{aligned}
g_{\mu \nu} &= -c^2 T_{\mu}T_{\nu}+ \Pi_{\mu \nu},&
g^{\mu \nu} &= -\tfrac{1}{c^2} V^{\mu}V^{\nu}+ \Pi^{\mu \nu}. 
\end{aligned}
\end{equation}
The metric, its inverse, and the spatial tensors can be written in terms of vielbeins as
\begin{equation} \label{eq:4}
\begin{aligned}
    g_{\mu \nu} &= \eta_{AB}E^A_{\mu}E^B_{\nu},
&
    g^{\mu \nu} &= \eta^{AB}\Theta_A^{\mu}\Theta_B^{\nu},
&
     \Pi_{\mu \nu} &= \eta_{ab}E^a_{\mu}E^b_{\nu},
&
     \Pi^{\mu \nu} &= \eta^{ab}\Theta_a^{\mu}\Theta_b^{\nu},
     \end{aligned}
\end{equation}
where $E^A_{\mu}$ and $\Theta_A^{\mu}$ are the vielbeins. Indices $A,B,\dots$ are vielbein labels running from 1 to $d+1$ (the dimension of $TM$) while $a,b,\dots$ are vielbein labels running from 1 to $d$ (the dimension of the $\mathrm{Hor}M$). Comparing the PUL parametrization with the vielbein definition we get ${E^{A}_{\mu}=(cT_{\mu},E^a_{\mu})}$ and ${\Theta^{\mu}_{A}=(-c^{-1}V^{\mu},\Theta^{\mu}_{a})}$. 

Following \cite{Hansen:2021fxi}, we assume that all fields are analytic in $c^2$ and expand them as follows:
\begin{equation}\label{eq:5}
\begin{aligned}
    V^{\mu} &=v^{\mu}+c^2 M^{\mu}+ O(c^4),
&
    T_{\mu} &=\tau_{\mu}+c^2 N_{\mu}+ O(c^4),
&
    \Theta_a^{\mu} &=\theta_a^{\mu}+c^2 \pi_a^{\mu}+ O(c^4),
\\
    E^a_{\mu} &=e^a_{\mu}+c^2 F^a_{\mu}+ O(c^4),
&
    \Pi^{\mu\nu} &=h^{\mu\nu}+c^2 \Phi^{\mu \nu}+ O(c^4),
&
     \Pi_{\mu\nu} &=h_{\mu\nu}+c^2 \Phi_{\mu \nu}+ O(c^4),
     \end{aligned}
\end{equation}
where $v^{\mu},M^{\mu},\tau_{\mu},N_{\nu},\theta_a^{\mu},\pi_a^{\mu},e^a_{\mu},F^a_{\mu},h^{\mu\nu},\Phi^{\mu\nu}$ are fields used to define geometries in the Carrollian limit. These fields are not all independent but they are related by two constraints. Thus, we can write $\tau_{\mu}$ and $\theta_a^{\mu}$ in terms of the other fields. Including more orders in $c^2$ leads to defining more fields that interpolate between the Carrollian theory (LO in the expansion) and the full theory on the manifold. Expanding the first equation in \eqref{eq:1}, we get
\begin{equation}\label{eq:6}
    \tau_{\mu}v^{\mu} + c^2 (\tau_{\mu}M^{\mu}+N_{\mu}v^{\mu})+c^4 N_{\mu}M^{\mu}=-1.
\end{equation}
Comparing the LO and NLO terms we arrive at
 \begin{equation}\label{eq:7}
 \begin{aligned}
     \tau_{\mu}v^{\mu} &=-1,&
     \tau_{\mu}M^{\mu}+N_{\mu}v^{\mu} &=0.
      \end{aligned}
 \end{equation}
Similarly, if we expand the second equation in \eqref{eq:1}, we obtain
\begin{equation}\label{eq:8}
     -\tau_{\nu}v^{\mu}+ h^{\mu \rho}h_{\rho \nu}+c^2 (h^{\mu \rho}\Phi_{\rho \nu}+ \Phi^{\mu\rho}h_{\rho \nu}- M^{\mu} \tau_{\nu} - v^{\mu}N_{\nu}) + c^4 \Phi^{\mu \rho}\Phi_{\rho \nu}= \delta^{\mu}_{\nu},
 \end{equation}
which by comparison of LO and NLO terms gives
 \begin{equation}\label{eq:9}
 \begin{aligned}
     -\tau_{\nu}v^{\mu}+ h^{\mu \rho}h_{\rho \nu} &= \delta^{\mu}_{\nu},&
h^{\mu \rho}\Phi_{\rho \nu}+ \Phi^{\mu}h_{\rho \nu}- M^{\mu} \tau_{\nu} - v^{\mu}N_{\nu} &=0.
 \end{aligned}
 \end{equation}
Now, by expanding \eqref{eq:3} we also get
\begin{equation}\label{eq:10}
    h_{\mu\nu} + c^2 \Phi_{\mu\nu}= \delta_{ab} e^a_{\mu}e^b_{\nu} + c^2 \delta_{ab}(F^a_{\mu}e^b_{\nu}+e^a_{\mu}F^b_{\nu}) + c^4 \delta_{ab}F^a_{\mu}F^b_{\nu}.
\end{equation}
and after comparing the LO and NLO terms, we arrive at
\begin{equation}\label{eq:11}
\begin{aligned}
    h_{\mu\nu} &=\delta_{ab}e^a_{\mu}e_{\nu}^b,
&
    \Phi_{\mu\nu} &=\delta_{ab}(F^a_{\mu}e^b_{\nu}+e^a_{\mu}F^b_{\nu}).
    \end{aligned}
\end{equation}
Similarly,
\begin{equation}\label{eq:12}
\begin{aligned}
    h^{\mu\nu} &=\delta^{ab}\theta_a^{\mu}\theta^{\nu}_b,
&
    \Phi^{\mu\nu} &=\delta^{ab}(\theta_a^{\mu}\pi^{\nu}_{b}+\pi_a^{\mu}\theta^{\nu}_{b}).
    \end{aligned}
\end{equation}
Remark that the induced metric $\boldsymbol{h}$ and the set of all $\boldsymbol{v}\in\mathcal{V}$, give rise to the Carrollian spacetime $(\mathcal{C},\mathcal{V},\boldsymbol{h})$.
%from Appendix~\ref{sec:geomstructure}, where $\mathcal{C}$ represents the limit of $M$.

To derive a compatible connection with the PUL parametrization \cite{Hansen:2021fxi,Ciambelli:2019lap}, we notice that $V^{\mu}$ and $\Pi_{\mu\nu}$ are invariant under Carroll boosts. Thus, they must be covariantly constant. Although this cannot determine a connection uniquely, it was argued in Appendix B of \cite{Hansen:2021fxi} that the most convenient choice is
\begin{equation}\label{eq:13}
C^{\rho}_{\mu\nu} = -V^{\rho}\partial_{(\mu}T_{\nu)}- V^{\rho}T_{(\mu}\pounds_{\boldsymbol{V}}T_{\nu)} + \tfrac{1}{2} \Pi^{\rho \lambda}\big[ \partial_{\mu} \Pi_{\nu \lambda}+ \partial_{\nu} \Pi_{\lambda \mu}- \partial_{\lambda}\Pi_{\mu \nu}\big]- \Pi^{\rho \lambda}T_{\nu}\mathcal{K}_{\mu \lambda},
\end{equation} 
where $\mathcal{K}_{\mu \lambda}= -\tfrac{1}{2} \pounds_{\boldsymbol{V}} \Pi_{\mu \lambda}$ is the extrinsic curvature (or the second fundamental from). 
The connection $C^{\rho}_{\mu\nu}$ has a nonzero torsion given by
\begin{equation}\label{eq:14}
    T^{\rho}_{\mu\nu}=2\Pi^{\rho\lambda}T_{[\mu}\mathcal{K}_{\nu]\lambda},
\end{equation}
which, to the LO, reads
\begin{equation}\label{torsion}
    T^{\rho}_{\mu\nu}=2h^{\rho\lambda}\tau_{[\mu}K_{\nu]\lambda}.
\end{equation}

To proceed parameterizing the Riemann tensor of the Levi-Civita connection, we write its Christoffel symbols $\Gamma^{\rho}_{\mu \nu}$ in terms of the PUL fields using \eqref{eq:3} and \eqref{eq:4}. The result is
\begin{equation}\label{eq:15}
\begin{aligned}
   & \Gamma^{\rho}_{\mu \nu} = \tfrac{1}{c^2}\big[-\tfrac{1}{2}V^{\rho}V^{\lambda}\partial_{\mu}\Pi_{\nu \lambda}- \tfrac{1}{2}V^{\rho} V^{\lambda} \partial_{\nu}\Pi_{\lambda \mu} + \tfrac{1}{2}V^{\rho}V^{\lambda} \partial_{\lambda}\Pi_{\mu \nu}\big] + \tfrac{1}{2}\big[\Pi^{\rho \lambda} \partial_{\mu}\Pi_{\nu \lambda}
    + \Pi^{\rho \lambda} \partial_{\nu}\Pi_{\lambda \mu} - \Pi^{\rho \lambda} \partial_{\lambda}\Pi_{\mu \nu}\\
    &\feq + V^{\rho}V^{\lambda} \partial_{\mu}(T_{\nu} T_{\lambda}) + 
    V^{\rho}V^{\lambda} \partial_{\nu}(T_{\mu} T_{\lambda})
    - V^{\rho}V^{\lambda} \partial_{\lambda}(T_{\nu} T_{\mu})\big]
    + c^2 \big[ \Pi^{\rho \lambda}\partial_{\mu}(T_{\nu}T_{\lambda})- \Pi^{\rho \lambda}\partial_{\nu}(T_{\mu}T_{\lambda}) + \Pi^{\rho \lambda}\partial_{\lambda}(T_{\nu}T_{\mu})\big]\;.
\end{aligned}    
\end{equation}
With the help of the coordinate expression of the Lie derivative we can rewrite $\Gamma^{\rho}_{\mu \nu}$ as
\begin{equation}\label{eq:16}
    \Gamma^{\rho}_{\mu \nu}= \tfrac{1}{c^2}\big[-V^{\rho} \mathcal{K}_{\mu \nu}\big] + \big[ C^{\rho}_{\mu\nu} + \Pi^{\rho \lambda}T_{\nu}\mathcal{K}_{\mu \lambda} \big] + c^2 \big[-T_{(\mu}\Pi^{\rho \lambda}B_{\nu)\lambda}\big], 
\end{equation}
where $B_{\mu\nu}=\partial_{\mu}T_{\nu}-\partial_{\nu}T_{\mu}$ is the exterior derivative of the covector $T_{\mu}$, which is the same as Eq. (2.21) in \cite{Hansen:2021fxi}. Finally, we are equipped to parameterize the Riemann tensor of $\Gamma^{\rho}_{\mu \nu}$,
\begin{equation}\label{eq:17}
    R^{\rho}{}_{\lambda \mu \nu}= \partial_{\mu}\Gamma^{\rho}_{\nu \lambda} - \partial_{\nu} \Gamma^{\rho}_{\mu \lambda} + \Gamma^{\rho}_{\mu \sigma} \Gamma^{\sigma}_{\nu \lambda} - \Gamma^{\rho}_{\nu \sigma}\Gamma^{\sigma}_{\mu \lambda}.
\end{equation}
Inserting \eqref{eq:15}, we obtain
\begin{equation}\label{eq:18}
\begin{aligned}
    &R^{\rho}{}_{\lambda \mu \nu} = \tfrac{1}{c^2}\big[2V^{\rho}\nabla_{[\nu}\mathcal{K}_{\mu]\lambda} + 2V^{\rho}\mathcal{K}_{\lambda\sigma}C^{\sigma}_{[\nu\mu]} + 2V^{\rho}T_{\lambda}\mathcal{K}^{\alpha}_{[\nu}\mathcal{K}_{\mu]\alpha}
     +2\mathcal{K}_{\lambda[\nu}\mathcal{K}^{\rho}_{\mu]} \big] 
    +\big[\overset{c}{R}{}^{\rho}{}_{\lambda\mu\nu}+ 2\nabla_{[\mu}(\mathcal{K}_{\nu]}^{\rho}T_{\lambda})+2C^{\sigma}_{[\mu\nu]}T_{\lambda}\mathcal{K}^{\rho}_{\sigma}\\ &\feq + V^{\rho}\mathcal{K}_{\mu \sigma}T_{(\nu}B_{\lambda )}{}^{\sigma}
    - V^{\rho}\mathcal{K}_{\nu \sigma}T_{(\mu}B_{\lambda )}{}^{\sigma} + T_{(\mu}B_{\sigma ) }{}^{\rho}V^{\sigma} \mathcal{K}_{\nu \lambda} - T_{(\nu}B_{\sigma )}{}^{\rho}V^{\sigma} \mathcal{K}_{\mu \lambda} \big] 
    +c^2\big[\nabla_{\nu}(T_{(\mu}B_{\lambda)}{}^{\rho})-\nabla_{\mu}(T_{(\nu}B_{\lambda)}{}^{\rho})\\ &\feq + 2C_{[\nu\mu]}^{\sigma}T_{(\sigma}B_{\lambda)}{}^{\rho}      ) 
     - T_{( \mu}B_{\sigma )}{}^{\rho}T_{\lambda}\mathcal{K}_{\nu}^{\sigma} + T_{( \nu}B_{\sigma )}{}^{\rho} T_{\lambda}\mathcal{K}_{\mu}{}^{\sigma}\big] 
    +c^4\big[T_{(\mu}B_{\sigma )}{}^{\rho}T_{( \nu}B_{\lambda )}{}^{\sigma} - T_{( \nu}B_{\sigma )}{}^{\rho}T_{( \mu}B_{\lambda )}{}^{\sigma}\big],
    \end{aligned}
\end{equation}
where $\overset{c}{R}{}^{\rho}{}_{\lambda\mu\nu}=\partial_{\mu}C^{\rho}_{\nu \lambda}-\partial_{\nu}C^{\rho}_{\mu \lambda}+ C^{\rho}_{\mu \sigma}C^{\sigma}_{\nu \lambda} - C^{\rho}_{\nu \sigma}C^{\sigma}_{\mu \lambda}$ and indices lowering/raising for $\mathcal{K}_{\mu\nu}$ and $B_{\mu\nu}$ was done by the induced metric and its inverse.

\section{Carrollian expansion of GR}\label{sec:4}

Having derived the PUL parametrization of the Riemann tensor in \eqref{eq:18}, we can now review the Carrollian expansion of the GR following \cite{Hansen:2021fxi}. Recall that the Einstein-Hilbert action in four dimensions ($d=3$) is,
\begin{equation}\label{eq:GRaction}
    S=c^3\int R \sqrt{-g}d^{4}x\;.
\end{equation}

Let us first calculate the PUL parametrization of the Ricci scalar $R$. By contracting $\rho$ and $\mu$ in \eqref{eq:18}, we can write the Ricci tensor in the form
\begin{equation}\label{eq:19}
    \begin{aligned}
       & R_{\lambda \nu} = \tfrac{1}{c^2}\big[-\nabla_{\mu}(V^{\mu}\mathcal{K}_{\nu \lambda}) - 2 V^{\mu}C^{\sigma}_{[\mu \lambda]}\mathcal{K}_{\nu \sigma}+ \mathcal{K}_{\nu \lambda} \mathcal{K} - \mathcal{K}_{\mu \lambda} \mathcal{K}_{\nu}^{\mu}\big]
        + \big[\overset{c}{R}_{\lambda \nu} + \nabla_{\mu}(T_{\lambda}\mathcal{K}_{\nu}^{\mu})- \nabla_{\nu}(T_{\lambda}\mathcal{K})
        +2 C^{\mu}_{[\nu \beta]}T_{\lambda}\mathcal{K}_{\mu}{}^{\beta} \\ &\feq + \mathcal{K}_{(\nu}^{\alpha}B_{\lambda) \alpha}  - \tfrac{1}{2} V^{\mu} \mathcal{K}_{\nu}^{\alpha}T_{\lambda}B_{\mu \alpha} - \tfrac{1}{2}T_{\nu}V^{\sigma} B_{\sigma \alpha} \mathcal{K}_{\lambda}^{\alpha}\big] 
        +c^2\big[-\nabla_{\mu}(T_{(\nu}B_{\lambda)}{}^{\mu}) +2 C^{\sigma}_{[\nu \mu]}T_{(\sigma}B_{\lambda)}{}^{\mu}
       + T_{(\nu}B_{\sigma)}{}^{\mu} T_{\lambda} \mathcal{K}_{\mu}{}^{\sigma} \big]
       \\ &\feq + c^4 \big[-\tfrac{1}{4}T_{\nu}T_{\lambda}B^{\mu\nu}B_{\mu \nu}\big],
    \end{aligned}
\end{equation}
where $\nabla_{\mu}$ is the covariant derivative corresponding to the connection $C^{\rho}_{\mu\nu}$. Here, we also introduced the trace of the extrinsic curvature, ${\mathcal{K}=\Pi^{\mu\nu}\mathcal{K}_{\mu\nu}}$, and the Ricci tensor of the connection $C^{\rho}_{\mu\nu}$, 
\begin{equation}
    \overset{c}{R}_{\lambda \nu}= \partial_{\mu}C^{\mu}_{\nu \lambda} - \partial_{\nu}C^{\mu}_{\mu \lambda} + C^{\mu}_{\mu \sigma}C^{\sigma}_{\nu \lambda} - C^{\mu}_{\nu \sigma}C^{\sigma}_{\mu \lambda}.
\end{equation}
The PUL parametrization of the Ricci scalar is obtained by contraction with the inverse metric and employing ${\Pi^{\lambda \nu}\nabla_{\mu}(V^{\mu}\mathcal{K}_{\nu\lambda}) = \nabla_{\nu}(V^{\nu}\mathcal{K})}$. The result is
\begin{equation}\label{eq:20}
    \begin{aligned}
        R &= \tfrac{1}{c^2}\big[\mathcal{K}^2 - \mathcal{K}_{\mu \nu} \mathcal{K}^{\mu\nu}
         - 2 \nabla_{\nu}(V^{\nu}\mathcal{K})\big]
        + \big[-\overset{c}{R} + \Pi^{\lambda \nu} \nabla_{\mu}(T_{\lambda} \mathcal{K}_{\nu}{}^{\mu})
        - \Pi^{\lambda \nu} \nabla_{\nu}(T_{\lambda} \mathcal{K})   +  V^{\lambda}V^{\nu}\nabla_{\mu}(T_{(\nu}  B_{\lambda)}{}^{\mu})\\ &\feq - V^{\lambda}V^{\nu}\nabla_{\nu}(T_{(\mu} B_{\lambda)}{}^{\mu})\big]
        + c^2\big[-\Pi^{\lambda \nu} \nabla_{\mu}(T_{(\nu} B_{\lambda)}{}^{\mu}) + \Pi^{\lambda \nu}\nabla_{\nu}(T_{(\mu}  B_{\lambda)}{}^{\mu}) - \tfrac{1}{4} B_{\mu\nu} B^{\mu \nu}\big],
    \end{aligned}
\end{equation}
where $\overset{c}{R}=\Pi^{\mu\nu}\overset{c}{R}_{\mu\nu}$. We used $V^{\mu}\overset{c}{R}_{\mu\nu}=0$ in the calculations.

Using the relation ${\nabla_{\rho}\Pi^{\mu\nu}=-V^{(\mu}\Pi^{\nu)\sigma}B_{\sigma \lambda}[\delta_{\rho}^{\lambda}-V^{\lambda}T_{\rho}]}$, we can find that ${\Pi^{\lambda \nu}\nabla_{\mu}(T_{\lambda} \mathcal{K}_{\nu}^{\mu})=0}$, ${\Pi^{\lambda \nu}\nabla_{\nu}(T_{\lambda}\mathcal{K})=0}$, ${V^{\lambda}V^{\nu} \nabla_{\mu}(T_{(\nu}B_{\lambda)}{}^{\mu})=-V^{\lambda}\nabla_{\mu}(B_{\lambda}^{\ \mu})}$ and ${-\Pi^{\lambda \nu}\nabla_{\mu}(T_{(\nu}B_{\lambda)}{}^{\mu})=\tfrac{1}{2}B^{\mu\nu}B_{\mu\nu}}$. Employing these identities, the Ricci scalar simplifies to
\begin{equation}\label{eq:21}
    \begin{aligned}
        R &= \tfrac{1}{c^2}\big[\mathcal{K}^2-\mathcal{K}^{\mu\nu}\mathcal{K}_{\mu\nu}-2\nabla_{\nu}(V^{\nu}\mathcal{K})\big]+\big[-\overset{c}{R}-\nabla_{\mu}(V^{\nu}B_{\nu}^{\ \mu})]
        +c^2[\tfrac{1}{4}B^{\mu\nu}B_{\mu\nu}\big].
    \end{aligned}
\end{equation}
 Furthermore, we can separate the total derivative terms as they corresponds to the boundary terms in actions of physical theories. Finally, the PUL parametrization of the Ricci scalar can be written in the form
\begin{equation}\label{eq:22}
    \begin{split}
        R= \tfrac{1}{c^2}\big[\mathcal{K}^2-  \mathcal{K}^{\mu \nu} \mathcal{K}_{\mu \nu} \big] + \big[- \overset{c}{R}\big]+ c^2\big[\tfrac{1}{4} B^{\mu\nu}B_{\mu \nu}\big] + \text{boundary terms},
    \end{split}
\end{equation}
where we collected all the boundary terms from all orders. Note that the boundary terms will be used in the calculation of quadratic curvature terms. (They are not important in this section since we compute the LO of GR.)

\begin{comment}

Hence, the PUL parameterization of the Einstein-Hilbert action is then
\begin{equation}\label{eq:28}
    S= c^2 \int \tfrac{1}{c^2}\big[\mathcal{K}^2-  \mathcal{K}^{\mu \nu} \mathcal{K}_{\mu \nu} \big] + \big[-\overset{c}{R}\big]+ c^2\big[ \tfrac{1}{4} B^{\mu\nu}B_{\mu \nu}\big] Ed^4x,
\end{equation}
where $E= \det(T_{\mu},E^a_{\mu})$, so after the Carrollian expansion to the LO, i.e. the electric limit, we arrive at
\end{comment}
Hence, the (electric) Carrollian limit of the GR action is
\begin{equation}\label{eq:CarrolActionGR}
    S= c^2 \int \big[K^2-  K^{\mu \nu} K_{\mu \nu}\big] ed^4x,
\end{equation}
where $K_{\mu\nu}=-\tfrac{1}{2}\pounds_{\boldsymbol{v}} h_{\mu\nu}$ and $e=\det(\tau_{\mu},e^a_{\mu})$.

Varying this action we get the constraints
\begin{equation}\label{GR constraints}
\begin{aligned}
    K^2-  K^{\mu \nu} K_{\mu \nu} &=0,
    &
    h^{\nu \alpha} \nabla_{\alpha}\big[K_{\mu\nu}-Kh_{\mu\nu}\big] &=0.
\end{aligned}
\end{equation}
and the evolution equation
\begin{equation}\label{eq:34}
    \pounds_{\boldsymbol{v}} K_{\mu\nu}= -2 K_{\mu}^{\alpha}K_{\nu \alpha} + KK_{\mu\nu}.
\end{equation}

\section{Carrollian expansion of quadratic gravity}\label{sec:5}

Quadratic gravity is a theory where quadratic curvature terms are added to the action, which makes it renormalizable \cite{Stelle:1976gc,Stelle:1977ry}. It also emerges from string theory by imposing a cutoff for the maximum possible momenta \cite{Nenmeli:2021orl}. The action for the theory is given by
\begin{equation}\label{eq:35}
    S= c^3\int   \big[R-\alpha R^{\mu \nu}R_{\mu\nu} + \beta R^2\big]\sqrt{-g}d^4x.
\end{equation}

In Sec.~\ref{sec:4} we computed the PUL parametrization of $R$. Now, we will do the same also for the two other terms in the action, $R^{\mu \nu}R_{\mu\nu}$ and $R^2$. Using \eqref{eq:19}, we can find the PUL parametrization of $R^{\mu \nu}R_{\mu\nu}$,

\begin{equation}\label{eq:36}
    R^{\mu \nu}R_{\mu\nu}=\tfrac{1}{c^4}\overset{(-4)}{R^{\mu \nu}R_{\mu\nu}} + \tfrac{1}{c^2}\overset{(-2)}{R^{\mu \nu}R_{\mu\nu}} + \overset{(0)}{R^{\mu \nu}R_{\mu\nu}} + c^4\overset{(4)}{R^{\mu \nu}R_{\mu\nu}}
\end{equation}
where
\begin{equation}
    \begin{aligned}
    \overset{(-4)}{R^{\mu \nu}R_{\mu\nu}}&=\Pi^{\nu\alpha}\Pi^{\lambda\beta} \nabla_{\mu}(V^{\mu}\mathcal{K}_{\alpha \beta})\nabla_{\rho}(V^{\rho}\mathcal{K}_{\nu \lambda}) -2 \mathcal{K}^{\alpha\beta}\mathcal{K}\nabla_{\mu}(V^{\mu}\mathcal{K}_{\alpha \beta}) + \mathcal{K}^{\lambda \nu}\mathcal{K}_{\lambda \nu} \mathcal{K}^2 
    - V^{\mu}V^{\nu}\nabla_{\mu}\mathcal{K}\nabla_{\nu}\mathcal{K}\\ &\feq + 2 \mathcal{K}_{\alpha\beta} \mathcal{K}^{\alpha \beta} V^{\nu}\nabla_{\nu}\mathcal{K}-(\mathcal{K}^{\mu\nu}\mathcal{K}_{\mu\nu})^2, \\
\overset{(-2)}{R^{\mu \nu}R_{\mu\nu}}&= -2\overset{c}{R}{}^{\lambda \nu} \nabla_{\mu}(V^{\mu}\mathcal{K}_{\lambda \nu})- \nabla_{\mu}(V^{\mu}\mathcal{K}_{\alpha\beta})\mathcal{K}^{\rho \beta} B^{\alpha}{}_{\rho} - \tfrac{1}{2} \mathcal{K}^{\rho \alpha} B^{\beta}{}_{ \rho}V^{\mu}\nabla_{\mu}\mathcal{K}_{\alpha\beta}  + 2\overset{c}{R}_{\lambda\nu}\mathcal{K}^{\lambda \nu} \mathcal{K}
    + \mathcal{K}^{\lambda \nu} \mathcal{K} \mathcal{K}^{\alpha}_{\lambda} B_{\nu \alpha}\\ &\feq - \Pi^{\nu\alpha}\nabla_{\mu}\mathcal{K}^{\mu}_{\alpha}\nabla_{\rho}\mathcal{K}^{\rho}_{\nu} + 2\Pi^{\nu\alpha}\nabla_{\mu}\mathcal{K}^{\mu}_{\alpha}\nabla_{\nu}\mathcal{K} + V^{\lambda} \nabla_{\mu}\mathcal{K}^{\mu}_{\alpha} \mathcal{K}^{\rho \alpha} B_{\lambda \rho} 
    - \Pi^{\nu\alpha}\nabla_{\nu}\mathcal{K}\nabla_{\alpha}\mathcal{K} -V^{\lambda}\nabla_{\alpha}\mathcal{K}\mathcal{K}^{\alpha \epsilon} B_{\lambda \epsilon} \\ &\feq - 2V^{\nu}V^{\alpha} \nabla_{\mu}(B_{\alpha}{}^{\mu})\nabla_{\nu}\mathcal{K}
    - V^{\alpha}\mathcal{K}^{\sigma \lambda} B_{\sigma\alpha} V^{\rho} \mathcal{K}^{\beta}_{\lambda} B_{\rho \beta} + 2 V^{\lambda} \nabla_{\mu}((dT)_{\lambda}{}^{\mu}) \mathcal{K}^{\alpha\beta}\mathcal{K}_{\alpha\beta},    \\
    \overset{(0)}{R^{\mu \nu}R_{\mu\nu}} &= \tfrac{1}{2} \mathcal{K}^{\alpha\beta}\mathcal{K}_{\alpha\beta}B^{\mu\nu}B_{\mu\nu} - \Pi^{\nu\alpha}\nabla_{\alpha}\mathcal{K}\nabla_{\rho}(B_{\nu}{}^{\rho}) - \tfrac{1}{4} V^{\sigma} \mathcal{K}^{\nu\rho} B_{\sigma\rho} \nabla_{\mu}(B_{\nu}{}^{\mu})
     - \tfrac{1}{2} V^{\lambda} V^{\alpha} \nabla_{\mu}(B_{\alpha}{}^{\mu})\nabla_{\nu}(B_{\lambda}{}^{\nu})\\ &\feq + \tfrac{3}{2}\overset{c}{R}{}^{\alpha\lambda}\mathcal{K}^{\beta}_{\lambda}B_{\alpha\beta}
     + \overset{c}{R}{}^{\mu\nu}\overset{c}{R}_{\mu\nu} + \tfrac{1}{4} \mathcal{K}_{\beta\lambda}\mathcal{K}^{\rho\lambda}B^{\nu\beta}B_{\nu\rho} + \tfrac{1}{4}\mathcal{K}^{\beta\lambda}\mathcal{K}^{\alpha\rho}B_{\alpha\beta}B_{\lambda\rho}, \\
     \overset{(4)}{R^{\mu \nu}R_{\mu\nu}} &= \tfrac{1}{16}(B^{\alpha\beta}B_{\alpha\beta})^2 .  
    \end{aligned}
\end{equation}
By expanding this expression to the LO, we arrive at
\begin{equation}\label{eq:37}
\begin{aligned}
     R^{\mu \nu}R_{\mu\nu} &= \tfrac{1}{c^4}\big[h^{\nu\alpha}h^{\lambda\beta} \nabla_{\mu}(v^{\mu}K_{\alpha \beta})\nabla_{\rho}(v^{\rho}K_{\nu \lambda}) -2 K^{\alpha\beta} K \nabla_{\mu}(v^{\mu} K_{\alpha \beta}) + K^{\lambda \nu} K_{\lambda \nu} K^2 \\ &\feq
    - v^{\lambda}v^{\nu}\nabla_{\mu}K\nabla_{\nu}K + 2 K_{\alpha\beta} K^{\alpha \beta} v^{\nu}\nabla_{\nu}K-(K^{\mu\nu}K_{\mu\nu})^2\big].
    \end{aligned}
\end{equation}
The PUL parametrization of $R^2$ can be computed from \eqref{eq:20}, 
\begin{equation}\label{eq:38}
    \begin{aligned}
        &R^2 = \tfrac{1}{c^4}\big[\mathcal{K}^4 - 2\mathcal{K}^2\mathcal{K}^{\mu\nu}\mathcal{K}_{\mu\nu} - 4\mathcal{K}^2\nabla_{\nu}(V^{\nu}\mathcal{K})+ (\mathcal{K}^{\mu\nu}\mathcal{K}_{\mu\nu})^2 + 4\mathcal{K}^{\mu\nu}\mathcal{K}_{\mu\nu}\nabla_{\alpha}(V^{\alpha}\mathcal{K}) + 4\nabla_{\mu}(V^{\mu}\mathcal{K})\nabla_{\nu}(V^{\nu}\mathcal{K})\big] \\ &\feq
        + \tfrac{1}{c^2}\big[-\mathcal{K}^2\overset{c}{R}  - \mathcal{K}^2\nabla_{\mu}(V^{\lambda}B_{\lambda}^{\ \mu}) + \mathcal{K}_{\mu\nu}\mathcal{K}^{\mu\nu}\overset{c}{R} + \mathcal{K}_{\mu\nu}\mathcal{K}^{\mu\nu}\nabla_{\rho}(V^{\lambda}B_{\lambda}^{\ \rho}) + 2\overset{c}{R}\nabla_{\mu}(V^{\mu}\mathcal{K}) 
        + 2\nabla_{\mu}(V^{\mu}\mathcal{K})\nabla_{\nu}(V^{\lambda}B_{\lambda}^{\ \nu})\big] \\ &\feq + \big[\tfrac{1}{2}\mathcal{K}B^{\mu\nu}B_{\mu\nu}+ \big(\overset{c}{R}\big)^2 - \tfrac{1}{2}\mathcal{K}^{\mu\nu}\mathcal{K}_{\mu\nu}B^{\sigma \rho}B_{\sigma \rho} - B^{\mu\nu}B_{\mu\nu} \nabla_{\rho}(V^{\rho}\mathcal{K}) 
         + \nabla_{\mu}(V^{\lambda}B_{\lambda}^{\ \mu})\nabla_{\rho}(V^{\sigma}B_{\sigma}^{\ \rho})\\ &\feq + 2\overset{c}{R} \nabla_{\mu}(V^{\lambda}B_{\lambda}^{\ \mu})\big]  + c^2\big[-\tfrac{1}{4} \overset{c}{R} B^{\mu\nu}B_{\mu\nu} 
        - B^{\mu\nu}B_{\mu\nu} \nabla_{\rho}(V^{\sigma}B_{\sigma}^{\ \rho})\big]  + c^4\big[\tfrac{1}{16}(B_{\mu\nu}B^{\mu\nu})^2\big],
    \end{aligned}
\end{equation}
and its Carrollian expansion to the LO is
\begin{equation}\label{eq:39}
    \begin{split}
        R^2 &= \tfrac{1}{c^4}\big[K^4 - 2K^2K^{\mu\nu}K_{\mu\nu} - 4K^2\nabla_{\nu}(v^{\nu}K)+ (K^{\mu\nu}K_{\mu\nu})^2 + 4K^{\mu\nu}K_{\mu\nu}\nabla_{\alpha}(v^{\alpha}K) + 4\nabla_{\mu}(v^{\mu}K)\nabla_{\nu}(v^{\nu}K)\big].
    \end{split}
\end{equation}

Substituting \eqref{eq:22}, \eqref{eq:37}, \eqref{eq:39} into the action \eqref{eq:35} we get
\begin{equation}\label{eq:40}
    \begin{aligned}
        &S =  \int  \big\{c^2\big[K^2- K^{\mu \nu} K_{\mu \nu}\big]- \alpha \big[h^{\nu\alpha}h^{\lambda\beta} \nabla_{\mu}(v^{\mu}K_{\alpha \beta})\nabla_{\rho}(v^{\rho}K_{\nu \lambda}) -2 K^{\alpha\beta} K \nabla_{\mu}(v^{\mu} K_{\alpha \beta}) + K^{\lambda \nu} K_{\lambda \nu} K^2 \\&\feq
    - v^{\mu}v^{\nu}\nabla_{\mu}K\nabla_{\nu}K  + 2 K_{\alpha\beta} K^{\alpha \beta} v^{\nu}\nabla_{\nu}K-(K^{\mu\nu}K_{\mu\nu})^2\big]+ \beta \big[K^4 - 2K^2K^{\mu\nu}K_{\mu\nu} 
    - 4K^2\nabla_{\nu}(v^{\nu}K) \\&\feq + (K^{\mu\nu}K_{\mu\nu})^2  + 4K^{\mu\nu}K_{\mu\nu}\nabla_{\rho}(v^{\rho}K) + 4\nabla_{\mu}(v^{\mu}K)\nabla_{\nu}(v^{\nu}K)\big]\big\} e d^4 x  .
    \end{aligned}
\end{equation}
Interestingly, this formula can be rewritten purely the extrinsic curvature $K_{\mu\nu}$ and its Lie derivatives along $\boldsymbol{v}$. In order to do that, we first write
\begin{equation}
    \pounds_{\boldsymbol{v}}K_{\mu\nu}=v^{\sigma}\nabla_{\sigma}K_{\mu\nu}+K_{\sigma\nu}\nabla_{\mu}v^{\sigma}+K_{\sigma\mu}\nabla_{\nu}v^{\sigma} - K_{\mu\nu}\nabla_{\sigma}v^{\sigma}+v^{\sigma}T^{\rho}_{\sigma\mu}K_{\rho\nu} + v^{\sigma}T^{\rho}_{\sigma\nu}K_{\rho\mu},
\end{equation}
where $T^{\rho}_{\mu\nu}$ is the torsion of the connection defined in Sec.~\ref{sec:2}. Since the PUL-parameterization vector $v^{\mu}$ is covariantly constant by definition and $T^{\rho}_{\mu\nu}$ is given by \eqref{torsion}, the relation reduces to
\begin{equation}\label{derivative relation}
    \pounds_{\boldsymbol{v}} K_{\mu\nu}=v^{\sigma}\nabla_{\sigma}K_{\mu\nu}- K^{\sigma}_{(\mu}K_{\nu)\sigma}.
\end{equation}
Substituting in \eqref{eq:40} and using the fact that $v^{\sigma}\nabla_{\sigma}$ acts on scalars simply as $\pounds_{\boldsymbol{v}}$, we get
\begin{equation}\label{lie der action}
    \begin{aligned}
      &S  =  \int \big\{c^2\big[K^2- K^{\mu \nu} K_{\mu \nu}\big]- \alpha  \big[   h^{\nu\alpha}h^{\lambda\beta}\pounds_{\boldsymbol{v}}K_{\alpha\beta}\pounds_{\boldsymbol{v}}K_{\nu\lambda} + 2 \pounds_{\boldsymbol{v}}K_{\nu\lambda}K^{\sigma(\nu}K^{\lambda)}_{\sigma} + K^{\sigma}_{(\alpha}K_{\beta)\sigma}K^{\rho(\alpha}K^{\beta)}_{\rho} \\ 
      &\phantom{=}  - (K_{\mu\nu}K^{\mu\nu})^2-2K^{\alpha\beta}K\pounds_{\boldsymbol{v}}K_{\alpha\beta}-2K^{\alpha\beta}KK^{\sigma}_{(\alpha}K_{\beta)\sigma}+ K^2K^{\mu\nu}K_{\mu\nu}- (\pounds_{\boldsymbol{v}} K)^2 +2K_{\mu\nu}K^{\mu\nu}\pounds_{\boldsymbol{v}} K  \big]\\
 &\phantom{=}  + \beta \big[  
 K^4 - 2K^2K_{\mu\nu}K^{\mu\nu}  - 4K^2\pounds_{\boldsymbol{v}} K + (K_{\mu\nu}K^{\mu\nu})^2 + 4K_{\mu\nu}K^{\mu\nu} \pounds_{\boldsymbol{v}} K + (\pounds_{\boldsymbol{v}} K)^2\big]
      \big\}ed^4 x   .
    \end{aligned}
\end{equation}

Note that only the first two terms have the factor $c^2$. Thus, assuming $\alpha$ and $\beta$ being independent of $c$, the Carrollian limit of the theory would exclude the first two terms coming from the Carrollian limit of the Ricci scalar. This means that the resulting theory would not couple to $R$ and it will be drastically different from the Carrollian limit of GR [cf. \eqref{eq:CarrolActionGR}]. Hence, $\alpha$ and $\beta$ should depend on $c$. In this case, we get an infinite number of nonequivalent Carrollian theories, but only four of them modify GR to LO or NLO. Notice that this limit is, as expected, ultralocal since there are no space derivatives in the Lagrangian, and therefore, there would not be space derivatives in the field equations. This means that the evolution of a point cannot be affected by neighboring points no matter how close they are.

Similar calculations were done in \cite{Niedermaier:2023hgk} by rescaling specific terms in the action. However, our approach gives more freedom to rescale terms differently and gives more nonequivalent theories. Other papers considered specific solutions for $f(R)$ gravity \cite{Abebe:2014hka,Abebe:2016obi,abebe2015irrotational}. A general classification of theories for the most general quadratic gravity theory will be provided in the next section.

 \section{Theories from the Carrollian limit of quadratic gravity}\label{sec:6}

In this section, we study Carrollian theories resulting from the Carrollian limit of quadratic gravity. Different (nonequivalent) theories arise from assuming different dependencies of $\alpha$ and $\beta$ on the speed of light $c$ in \eqref{lie der action}. Thus, we classify them as such and denote them by $(n,m)$, where $\alpha = c^n\alpha'$  and $\beta = c^m\beta'$. The relevant theories are listed in Tab.~\ref{table:1}. As mentioned above, not all theories are modifications to GR. For example, the theories with negative powers of $c$ in $\alpha$ or $\beta$ but also (0,0), (0,2) and (2,0) are not physically interesting since they are drastically different from GR at LO. It is easy to see that dependencies with odd powers of $c$ ultimately converge to one of the theories in Tab. \ref{table:1}. Theories with higher-power dependencies on $c$ cannot modify GR to the LO nor the NLO but to higher orders however, since $\alpha$ and $\beta$ dependencies on $c$ is a non perturbative assumption, having higher powers of $c$ in the action without being an overall factor can lead to inconsistencies in the Galilean limit. Thus, in what follows, we focus only on the the four interesting Carrollian theories (2,2), (2,4), (4,2), and (4,4).
 
\setlength{\arrayrulewidth}{0.5mm}
\renewcommand{\arraystretch}{2}
\setlength{\tabcolsep}{10pt}
\begin{table}
 \begin{tabular}{ |p{0.9cm}|p{6.6cm}|p{7cm}|  }
 \hline
 \multicolumn{3}{|c|}{Carrollian theories from quadratic gravity} \\
 \hline
 Theory  & Action contributing to the LO & Type of modification to the Carrollian limit of GR\\
 \hline
 (0,0) & $ S= c^3 \int  \big[-\alpha R^{\mu\nu}R_{\mu\nu}+\beta R^2\big] \sqrt{-g}d^4 x  $ & \textit{Not a modification of GR} \\
 \hline
 (0,2)   & $S= c^3 \int -\alpha R^{\mu\nu}R_{\mu\nu} \sqrt{-g}d^4 x$ & \textit{Not a modification of GR}\\
 \hline
 (2,0) & $ S= c^3 \int \beta R^2 \sqrt{-g}d^4 x $ & \textit{Not a modification of GR} \\
 \hline
 (2,2)    & $S= c^3 \int \big[R-\alpha R^{\mu \nu}R_{\mu\nu} + \beta R^2\big] \sqrt{-g}d^4 x$ & \textit{Modifies GR to the LO}\\
 \hline
 (2,4) &  $S= c^3 \int \big[R-\alpha R^{\mu \nu}R_{\mu\nu} \big]\sqrt{-g}d^4 x$   & \textit{Modifies GR to the LO with $R^{\mu\nu}R_{\mu\nu}$ terms and the NLO by $R^2$ terms}\\
 \hline
 (4,2) & $S= c^3 \int \big[R + \beta R^2\big] \sqrt{-g}d^4 x$  & \textit{Modifies GR to the LO with $R^2$ terms and the NLO by $R^{\mu\nu}R_{\mu\nu}$ terms} \\
 \hline
 (4,4) & $S= c^3\int R\sqrt{-g}d^4 x$ & \textit{Modifies GR in the NLO}\\
 \hline
\end{tabular}
\caption{This table summarizes some possible Carrollian theories arising from quadratic gravity that couple to $R$ at most in the NLO. We list the theories with factors of $c$ with non-negative powers since negative $c$ dependencies are clearly not modifications of the Carrollian limit of GR. For example, although (0,0) cannot be a modification to the Carrollian limit of GR, we can say that $R$ terms are a NLO modification of this theory. There are other geometries which are modifications to the listed geometries like (0,4) which can be regarded as a next-to-next-to-leading order modification of (0,2) while GR itself is the NLO. We can extend the list indefinitely adding more geometries modifying GR to higher orders but here we focus on the LO and NLO.}
\label{table:1}
\end{table}

\subsection{(2,2) Carrollian theory}

Consider the case where $\alpha$ and $\beta$ are quadratic in the speed of light, ${\alpha = c^2 \alpha'}$, ${\beta = c^2\beta'}$, with $\alpha'$ and $\beta'$ being constants independent of $c$. We will study the resulting action to the LO, i.e., the electric limit. From Tab.~\ref{table:1}, the action is
\begin{equation}\label{eq:41}
    S=  c^3\int \big[R-\alpha R^{\mu \nu}R_{\mu\nu} + \beta R^2\big] \sqrt{-g}d^4x.
\end{equation}
Writing $\alpha= c^2 \alpha'$ and $\beta = c^2 \beta'$, where $\alpha'$ and $\beta'$ are $c$ independent constants, we can write the action as
\begin{equation}\label{eq:42}
 S= \int  c^3 \big[R-c^2\alpha' R^{\mu \nu}R_{\mu\nu} + c^2 \beta' R^2\big]\sqrt{-g}d^4x, 
\end{equation}
which in the LO of the Carrollian expansion gives
\begin{equation}\label{eq:43}
\begin{aligned}
S =  c^2\int    &\big\{ \big[K^2- K^{\mu \nu} K_{\mu \nu}\big]- \alpha'  \big[   h^{\nu\alpha}h^{\lambda\beta}\pounds_{\boldsymbol{v}}K_{\alpha\beta} \pounds_{\boldsymbol{v}}K_{\nu\lambda} + 2 \pounds_{\boldsymbol{v}}K_{\nu\lambda}K^{\sigma(\nu}K^{\lambda)}_{\sigma} + K^{\sigma}_{(\alpha}K_{\beta)\sigma}K^{\rho(\alpha}K^{\beta)}_{\rho} -2K^{\alpha\beta}K\pounds_{\boldsymbol{v}}K_{\alpha\beta}\\ 
      &\phantom{=}-2K^{\alpha\beta}KK^{\sigma}_{(\alpha}K_{\beta)\sigma}+ K^2K^{\mu\nu}K_{\mu\nu} - (\pounds_{\boldsymbol{v}}K)^2 +2K_{\mu\nu}K^{\mu\nu}\pounds_{\boldsymbol{v}}K - (K_{\mu\nu}K^{\mu\nu})^2 \big] + \beta' \big[  
 K^4 - 2K^2K_{\mu\nu}K^{\mu\nu} \\
 &\phantom{=} - 4K^2\pounds_{\boldsymbol{v}}K + (K_{\mu\nu}K^{\mu\nu})^2 + 4K_{\mu\nu}K^{\mu\nu} \pounds_{\boldsymbol{v}}K + (\pounds_{\boldsymbol{v}}K)^2\big]
      \big\} ed^4 x
\end{aligned}
\end{equation}

Since the Carrollian expansion and the weak-field regime are not conflicting, the conditions to find tachyons remain the same. In \cite{Stelle:1977ry} it was found that the additional degrees of freedom have masses of\footnote{Remark that $\alpha$ and $\beta$ in our convention have opposite signs to the convention used in \cite{Stelle:1977ry}.} 
\begin{equation}\label{eq:44}
    \begin{aligned}
         m_0 &=\frac{1}{\sqrt{2}}\frac{1}{\sqrt{-\alpha}},& 
        m_2 &= \frac{1}{\sqrt{2}} \frac{1}{\sqrt{\alpha-3\beta}}.
    \end{aligned}
\end{equation}
The conditions to avoid tachyons are (at any order of the Carrollian expansion)
\begin{equation}\label{tachyonremoving}
        \alpha \leq 0, \quad
        \alpha -3\beta \geq 0,
\end{equation}
which translates to
\begin{equation}\label{eq:45}
        \alpha' \leq 0, \quad
        \alpha' -3\beta' \geq 0,
\end{equation}
in the case of (2,2) theory.

\subsection{(2,4) Carrollian theory}

Let us now investigate the case where ${\alpha = c^2 \alpha'}$ and ${\beta = c^4\beta'}$. The action is
\begin{equation}\label{eq:46}
    S= c^3\int \big[R-c^2\alpha' R^{\mu \nu}R_{\mu\nu} + c^4 \beta' R^2\big]\sqrt{-g}d^4x.
\end{equation}
To the LO in the Carrollian expansion, we get the action
\begin{equation}\label{eq:47}
\begin{aligned}
    S &= c^2 \int  \big\{\big[K^2-K^{\mu\nu}K_{\mu\nu}\big] - \alpha' \big[   h^{\nu\alpha}h^{\lambda\beta}\pounds_{\boldsymbol{v}}K_{\alpha\beta}\pounds_{\boldsymbol{v}}K_{\nu\lambda} + 2 \pounds_{\boldsymbol{v}}K_{\nu\lambda}K^{\sigma(\nu}K^{\lambda)}_{\sigma} + K^{\sigma}_{(\alpha}K_{\beta)\sigma}K^{\rho(\alpha}K^{\beta)}_{\rho} \\ 
      &\phantom{=}-2K^{\alpha\beta}K\pounds_{\boldsymbol{v}}K_{\alpha\beta}-2K^{\alpha\beta}KK^{\sigma}_{(\alpha}K_{\beta)\sigma}+ K^2K^{\mu\nu}K_{\mu\nu} - (\pounds_{\boldsymbol{v}}K)^2 +2K_{\mu\nu}K^{\mu\nu}\pounds_{\boldsymbol{v}}K - (K_{\mu\nu}K^{\mu\nu})^2\big] \big\}e d^4x. 
\end{aligned}
\end{equation}
Notice that this theory is the same as the Carrollian limit of $R-\alpha R_{\mu\nu}R^{\mu\nu}$. The conditions \eqref{tachyonremoving} to the LO reduce to ${\alpha' = 0}$. Thus, to the LO, the theory without tachyons is the same as the Carrollian limit of GR.

Assuming $\alpha'$ and $\beta'$ to be of the same numerical order, the conditions to the LO and NLO respectively are 
\begin{equation}\label{eq:48}
     \alpha'=0, \ \quad
     \beta' \leq 0.
\end{equation}
Thus, the theory without tachyons to the NLO would be
\begin{equation} \label{(2,4)theory}
    S= c^3\int\big[ R_{NLO} 
+ c^4 \beta'(R^2)_{LO} \big] \sqrt{-g}d^4x,
\end{equation}
where $R_{NLO}$ is the Ricci scalar expanded to the NLO and $(R^2)_{LO}$ is the LO of the Carrollian expansion of $R^2$.

\subsection{(4,2) Carrollian theory}

Considering the dependencies are $\alpha = c^4\alpha'$ and $\beta = c^2 \beta'$, the action is
\begin{equation}\label{eq:49}
     S= c^3\int \big[R-c^4\alpha' R^{\mu \nu}R_{\mu\nu} + c^2 \beta' R^2\big]  \sqrt{-g}d^4x.
\end{equation}
The corresponding LO action reads
\begin{equation}\label{eq:50}
    \begin{aligned}
    S=c^2\int & \big[\big(K^2-K^{\mu\nu}K_{\mu\nu}\big)+ \beta' \big[  
 K^4 - 2K^2K_{\mu\nu}K^{\mu\nu} - 4K^2\pounds_{\boldsymbol{v}}K + (K_{\mu\nu}K^{\mu\nu})^2 + 4K_{\mu\nu}K^{\mu\nu} \pounds_{\boldsymbol{v}}K + (\pounds_{\boldsymbol{v}}K)^2\big]\big]e d^4x. 
    \end{aligned}
\end{equation}

In this case the conditions \eqref{tachyonremoving} then reduce to ${\beta' \leq 0}$. Expanding the conditions to the NLO we obtain
\begin{equation}
        \beta' \leq 0, \ \quad
        \alpha'=0.
\end{equation}
Hence, this theory is equivalent to the Carrollian limit of $R-\beta R^2$ theory to all orders with NLO action being the same as \eqref{(2,4)theory}.

\subsection{(4,4) Carrollian theory}

If we consider $\alpha = c^4\alpha'$ and $\beta = c^4 \beta'$, then the action reads
\begin{equation}\label{eq:52}
     S= c^3\int   \big[R-c^4\alpha' R^{\mu \nu}R_{\mu\nu} + c^4 \beta' R^2\big]\sqrt{-g}d^4x.
\end{equation}
For this theory, the LO action is the same as GR. At the NLO and higher orders it will receive corrections from both $R^2$ and $R_{\mu\nu}R^{\mu\nu}$ terms. The conditions \eqref{tachyonremoving} are the same as in the (2,2) case.

\section{The magnetic limit}\label{magnetic limit section}

In this section we study the magnetic limit of the theories (2,4) and (4,2) because these two theories are free from tachyons and ghosts if $\beta' \leq 0$. The magnetic limit is obtained by truncating the NLO action such that the resulting action is invariant under Carroll symmetries. In the case of quadratic gravity we have to truncate it the same way as GR, i.e., we have to put all the NLO fields to zero. It is well known that the NLO captures all the dynamics of the Carrollian limit \cite{Hansen:2021fxi}. Thus, the field equations from the magnetic limit leads to corrections to the dynamics in GR and even more solutions that are non existent in GR.

\subsection{The magnetic limit of (2,4)}

Imposing the truncation 
\begin{equation}\label{truncation}
    M^{\mu}=N_{\mu}=\Phi_{\mu\nu}=\Phi^{\mu\nu}=0,
\end{equation}
we get the LO and NLO of the terms of \eqref{(2,4)theory} to be
\begin{equation}\label{(2,4) constraints}
    \begin{aligned}
        R_{LO}&= K^2-K_{\mu\nu}K^{\mu\nu},&
        R_{NLO}&= -\overset{c}{R}, &
        (R^2)_{LO}&=( K^2-K_{\sigma\rho}K^{\sigma\rho})( K^2-K_{\mu\nu}K^{\mu\nu}+4\pounds_{\boldsymbol{v}} K)-4(\pounds_{\boldsymbol{v}}K)^2.
    \end{aligned}
\end{equation}
As shown in the previous section, the LO of this theory is identical to the LO of GR, i.e., the constraints and the evolution equation are the same as \eqref{GR constraints} and \eqref{eq:34}. In the NLO, the LO constraints and evolution equations must hold, so they serve as constraints to the NLO field equations. Thus, taking the trace of \eqref{eq:34} we get
\begin{equation}
    h^{\mu\nu}\pounds_{\boldsymbol{v}}K_{\mu\nu}=-2K_{\mu\nu}K^{\mu\nu}+K^2,
\end{equation}
then noting that
\begin{equation}
    \pounds_{\boldsymbol{v}}K=h^{\mu\nu}\pounds_{\boldsymbol{v}}K_{\mu\nu} + 2K_{\mu\nu}K^{\mu\nu},
\end{equation}
we get
\begin{equation}\label{eq:51}
\pounds_{\boldsymbol{v}}K=K^2.   
\end{equation}
Thus, the constraints on the magnetic action Lagrangian are
\begin{equation}
    \begin{aligned}
        K^2-K^{\mu\nu}K_{\mu\nu} &=0,&
         h^{\nu \alpha} \nabla_{\alpha}\big[K_{\mu\nu}-Kh_{\mu\nu}\big] &=0, &
        \pounds_{\boldsymbol{v}}K &= K^2.
    \end{aligned}
\end{equation}
Notice that this is not a general equation. It is valid only in (2,4) and the theories whose LO is identical to GR. 

Using the above relations, we can write the action for the magnetic limit of (2,4) as
\begin{equation}\label{magnetic limit of (2,4)}
\begin{aligned}
    S=- \int d^4x & e\big[-\overset{c}{R} + \beta'\big[( K^2-K_{\sigma\rho}K^{\sigma\rho})( K^2-K_{\mu\nu}K^{\mu\nu}  + 4\pounds_{\boldsymbol{v}} K)-4(\pounds_{\boldsymbol{v}}K)^2 \big]+ \\ &\phantom{=} \lambda_1(K^2-K_{\mu\nu}K^{\mu\nu})+ \beta'\lambda_2(\pounds_{\boldsymbol{v}}K-K^2)\big],
    \end{aligned}
\end{equation}
where $\lambda_1$ and $\lambda_2$ are Lagrange multipliers. As expected, the theory (2,4) modifies the magnetic limit of GR with quartic terms in the extrinsic curvature and imposes an additional constraint. It would be interesting to see how these terms modify the dynamics of different solutions of the field equations especially black holes. We expect that this theory has more solutions than the Carrollian limit of GR, namely those corresponding to the Schwarzchild-Bach black holes \cite{Podolsky:2019gro,Pravdova:2023nbo}. If this is the case then one should examine if some terms can be considered as a flux that is analogous to the magnetic field in \cite{Bicak:2023vxs}. Here, however, the flux would come from the theory instead of being turned on by hand. 

\subsection{The magnetic limit of (4,2)}

This case is more complicated than (2,4) since the LO is more involved than that of GR. We first study the constraints and the evolution equations for the LO, then move on to the NLO. To the LO, the action is 
\begin{equation}\label{(4,2)LO action}
S=\int d^4x e[(K^2-K_{\sigma\rho}K^{\sigma\rho})(1+\beta'[K^2-K^{\mu\nu}K_{\mu\nu}+4\pounds_{\boldsymbol{v}}K])- \beta'(\pounds_{\boldsymbol{v}} K)^2].
\end{equation}
Varying with respect to $v^{\mu}$ and $h^{\mu\nu}$, we get the constraints
\begin{subequations}\label{(4,2)constraints}
\begin{align}
    (K^2-K_{\sigma\rho}K^{\sigma\rho})(1+\beta'[K^2-K^{\mu\nu}K_{\mu\nu}+4\pounds_{\boldsymbol{v}}K])- \beta'(\pounds_{\boldsymbol{v}} K)^2=0,\\
    h^{\mu\rho}\nabla_\mu(K_{\rho\nu}-Kh_{\rho\nu}+2\beta'[K_{\rho\nu}(-3(K^2-K_{\alpha\beta}K^{\alpha\beta})+4\pounds_{\boldsymbol{v}}K)-Kh_{\rho\nu}(K^2-K_{\alpha\beta}K^{\alpha\beta}+2\pounds_{\boldsymbol{v}}K)])=0.
      \end{align}
\end{subequations}
Varying with respect to $h^{\mu\nu}$ and using the constraints, the evolution equation is
\begin{equation}\label{(4,2)evolution equation}
    \begin{aligned}   
    & \ \ \ 2(KK_{\mu\nu}-K_{\mu}^{\sigma}K_{\nu\sigma})(1+\beta'(2(K^2-K_{\alpha\beta}K^{\alpha\beta})+4\pounds_{\boldsymbol{v}}K)) + 2(2\beta'(K^2-K_{\alpha\beta}K^{\alpha\beta})-\beta'\pounds_{\boldsymbol{v}}K)(\pounds_{\boldsymbol{v}}K_{\mu\nu}-4K^{\sigma}_{\mu}K_{\sigma\nu}) \\ &\phantom{=}
    + \pounds_{\boldsymbol{v}}\big[ (Kh_{\mu\nu}-K_{\mu\nu})(1+\beta'(2(K^2-K_{\alpha\beta}K^{\alpha\beta})+4\pounds_{\boldsymbol{v}}K)) \big] - 8\beta'\pounds_{\boldsymbol{v}}\big[ K_{\mu\nu} (2(K^2-K_{\alpha\beta}K^{\alpha\beta})-\pounds_{\boldsymbol{v}}K)\big] \\ &\phantom{=}
    +2\beta'\pounds_{\boldsymbol{v}}\pounds_{\boldsymbol{v}}\big[ 2(K^2-K_{\alpha\beta}K^{\alpha\beta})-\pounds_{\boldsymbol{v}}K \big]=0.
    \end{aligned}
  \end{equation} 
As expected, setting ${\beta'=0}$, the equation reduces to the evolution equation of GR. The corrections to GR due to the $R^2$ term are quartic in the extrinsic curvature. 

After truncation the NLO action reads
\begin{equation}
    \begin{aligned}
        S =c^3\int & e\big[\overset{c}{R}+ \beta'(-K^2+K_{\mu\nu}K^{\mu\nu}+2\pounds_{\boldsymbol{v}}K)(\overset{c}{R}+\nabla_{\alpha}(v^{\lambda}b_{\lambda}^{\ \alpha})) \big]d^4x.
    \end{aligned}
\end{equation}
However, the equations for the LO must also, so we have to add \eqref{(4,2)constraints} to the Lagrangian as a constraint,
\begin{equation}\label{magnetic action for (4,2)}
    \begin{aligned}
        &S =c^3\int  e\big[\overset{c}{R}+ \beta'(-K^2+K_{\mu\nu}K^{\mu\nu}+2\pounds_{\boldsymbol{v}}K)(\overset{c}{R}+\nabla_{\alpha}(v^{\lambda}b_{\lambda}^{\ \alpha})) \\ &\phantom{=} 
        + \lambda((K^2-K_{\sigma\rho}K^{\sigma\rho})(1+\beta'[K^2-K^{\mu\nu}K_{\mu\nu}+4\pounds_{\boldsymbol{v}}K])- \beta'(\pounds_{\boldsymbol{v}} K)^2)\big]d^4x,
    \end{aligned}
\end{equation}
where $\lambda$ is a Lagrange multiplier. Notice that the field equations for this action must include \eqref{(4,2)evolution equation}.

Now, we study a special case of the above equations where we treat $\pounds_{\boldsymbol{v}} K$ as an independent variable. Varying the action with respect to $v^{\sigma}$, we get the equations
\begin{equation}\label{v equations}
    \begin{aligned}
        (K^2-K_{\sigma\rho}K^{\sigma\rho})(1+\beta'[K^2-K^{\mu\nu}K_{\mu\nu}+4\pounds_{\boldsymbol{v}}K])- \beta'(\pounds_{\boldsymbol{v}} K)^2=0, & \ \ \
        h^{\rho\sigma}\nabla_{\sigma}(Kh_{\rho\mu}-K_{\rho\mu})=0.
    \end{aligned}
\end{equation}
Varying the action with respect to $\pounds_{\boldsymbol{v}} K$ and assuming $\pounds_{\boldsymbol{v}} K \neq 0$, we get
\begin{equation}\label{K equation}
    \pounds_{\boldsymbol{v}} K = 2(K^2-K_{\mu\nu}K^{\mu\nu}).
\end{equation}
From \eqref{v equations} and \eqref{K equation}, we get the equations
\begin{equation}\label{eq of motion of (4,2)}
    \begin{aligned}
        \pounds_{\boldsymbol{v}} K = \tfrac{-2}{5\beta'}, \quad
        K^2-K_{\mu\nu}K^{\mu\nu}= \tfrac{-1}{5\beta'}.
    \end{aligned}
\end{equation}
Varying the action with respect to $h^{\mu\nu}$ and using \eqref{eq of motion of (4,2)} we get 
\begin{equation}\label{h equation}
    \pounds_{\boldsymbol{v}}K_{\mu\nu} = -2K_{\mu}^{\sigma}K_{\sigma \nu}+Kh_{\mu\nu}.
\end{equation}
Collecting the independent field equations we get the system
\begin{equation}\label{(4,2) special case}
\begin{aligned}
    \pounds_{\boldsymbol{v}} K &= \tfrac{-2}{5\beta'}, &
        K^2-K_{\mu\nu}K^{\mu\nu} &= \tfrac{-1}{5\beta'}, & 
         -2K_{\mu}^{\sigma}K_{\sigma \nu}+Kh_{\mu\nu} &= \pounds_{\boldsymbol{v}}K_{\mu\nu},&
        h^{\rho\sigma}\nabla_{\sigma}(Kh_{\rho\mu}-K_{\rho\mu})&=0.
\end{aligned}
\end{equation}
It turns out that this system solves \eqref{(4,2)constraints} and \eqref{(4,2)evolution equation}. Thus, the solutions to the system \eqref{(4,2) special case} are also solutions to the full (4,2) equations at LO. Notice that this system solves the full theory but the converse is not true. This means that a solution for \eqref{(4,2) special case} is a solution for the full theory but its set of solutions is only a subset of that of the full theory. It is also worth mentioning that this system cannot reproduce GR without a cosmological constant i.e. it is not valid for $\beta'=0$.

Notice that \eqref{(4,2) special case} is similar to equations (4.18) in \cite{Hansen:2021fxi}, which describe GR with a cosmological constant, except for the evolution equation is the same as GR without a cosmological constant. Modifications to the gravitational sector to reproduce a cosmological constant (without adding a cosmological constant term in the Lagrangian) were studied in $f(R)$ gravity \cite{delaCruz-Dombriz:2006kob}. Thus, we can interpret the effect of the $R^2$ term to be an effective cosmological constant with the value ${-1}/{(10\beta')}$. We will leave the solutions of this system of equations to future works. Now we use them as constraints to write the action for the magnetic limit.

For the special case where $\pounds_{\boldsymbol{v}}K$ is considered independent, the NLO action reads

\begin{equation}
    \begin{aligned}
        S =c^3\int & e\big[\overset{c}{R}+ \beta'(-K^2+K_{\mu\nu}K^{\mu\nu}+2\pounds_{\boldsymbol{v}}K)(\overset{c}{R}+\nabla_{\alpha}(v^{\lambda}b_{\lambda}^{\ \alpha}))  + \lambda_1(\pounds_{\boldsymbol{v}}K + \tfrac{2}{5\beta'}) + \lambda_2(K^2-K_{\mu\nu}K^{\mu\nu}+\tfrac{1}{5\beta'})\big]d^4x,
    \end{aligned}
\end{equation}
where  $\lambda_1$ and $\lambda_2$ are Lagrange multipliers, and $b_{\mu\nu}=\partial_{\mu}\tau_{\nu}-\partial_{\nu}\tau_{\mu}$.

It is clear that the action contains a cosmological term. This is a direct result of the emergence of an effective cosmological constant in the LO equations. Like the magnetic limit action of (2,4), this action modifies the magnetic limit of GR but with a non zero cosmological constant. Applying this to the general magnetic limit action \eqref{magnetic action for (4,2)} we conclude that it includes a cosmological term in addition to terms that can be interpreted as flux. Notice that 
magnetic limits are no longer ultralocal due to the presence of spatial derivatives of the metric in the form of the Ricci scalar and terms containing the covariant derivative of $b_{\mu\nu}$. This allows some dynamics that was absent in the electric limit.

\section{Conclusions} \label{sec:concl}

In the present paper, we studied the electric and magnetic Carrollian limits of quadratic gravity. We calculated the PUL parametrization of terms with quadratic curvature in the action. After the Carrollian expansion, we saw that such terms are of the order of ${c^{-4}}$ while the Ricci scalar term is only of the order of ${c^{-2}}$. From that, we concluded that the Carrollian limit of quadratic gravity requires $\alpha$ and $\beta$ to depend on $c$ in a particular way so that the resulting theory is a modification of GR. We classified different limits according to the dependencies of $\alpha$ and $\beta$ on $c$. For example, the three of them $(0,0)$ (no dependence on $c$), $(0,2)$, and $(2,0)$ are not GR modifications because to the LO only the terms of order ${c^{-4}}$ survive, i.e., only the quadratic terms in curvature but not the Ricci scalar. The only four theories that are modifications of GR (to the LO and NLO) are summarized in Tab.~\ref{table:2} together with the corresponding modifications.
\setlength{\arrayrulewidth}{0.5mm}
\renewcommand{\arraystretch}{2}
\setlength{\tabcolsep}{10pt}
\begin{table}
 \begin{tabular}{ |p{0.9cm}|p{5cm}|p{5.7cm}|p{1.7cm}|}
 \hline
 \multicolumn{4}{|c|}{Carrollian theories from quadratic gravity after removing tachyons} \\
 \hline
 Theory & Action contributing to the LO & Action contributing to the NLO & Conditions\\
 \hline
 (2,2)    & 
 $\begin{aligned}[t]
     S= c^3{\textstyle\int} \big[R_{LO}-c^2\alpha' (R^{\mu \nu}R_{\mu\nu})_{LO} \\ + c^2\beta' (R^2)_{LO}\big]\sqrt{-g}d^4x
     \end{aligned}$ & $\begin{aligned}[t] S= c^3{\textstyle\int} \big[R_{NLO} -c^2\alpha' (R^{\mu \nu}R_{\mu\nu})_{NLO}\\  + c^2\beta' (R^2)_{NLO}\big]\sqrt{-g}d^4x \end{aligned}$ & $\begin{aligned}[t] &\alpha' \leq 0, \\ &\alpha' -3\beta' \geq 0\end{aligned}$\\
 \hline
 (2,4) &  $\begin{aligned}[t] S= c^3{\textstyle\int} \big[R_{LO} \big]\sqrt{-g}d^4x\end{aligned}$   & $\begin{aligned}[t] S= c^3{\textstyle\int} \big[R_{NLO} + c^4\beta'(R^2)_{LO} \big]\sqrt{-g}d^4x \end{aligned}$ & $\begin{aligned}[t] \alpha'=0 \\ \beta' \leq 0\end{aligned}$\\
 \hline
 (4,2) & $\begin{aligned}[t] S=  c^3{\textstyle\int} \big[R_{LO} + c^2\beta' (R^2)_{LO}\big]\sqrt{-g}d^4x \end{aligned}$  & $\begin{aligned}[t] S= c^3{\textstyle\int} \big[R_{NLO} + c^4 \beta' (R^2)_{NLO}\big]\sqrt{-g}d^4x \end{aligned}$ & $\begin{aligned}[t] \alpha'=0 \\ \beta' \leq 0\end{aligned}$\\
 \hline
 (4,4) & $\begin{aligned}[t] S= c^3{\textstyle\int} \big[R_{LO}\big]\sqrt{-g}d^4x \end{aligned}$ & $\begin{aligned}[t] S= c^3{\textstyle\int} \big[R_{NLO} -c^4\alpha' (R^{\mu \nu}R_{\mu\nu})_{NLO} \\+ c^4\beta' (R^2)_{NLO}\big]\sqrt{-g}d^4x \end{aligned}$ & $\begin{aligned}[t] &\alpha' \leq 0, \\ &\alpha' -3\beta' \geq 0 \end{aligned}$\\
 \hline
\end{tabular}
\caption{After imposing the conditions to remove tachyons, the set of resulting theories consists either of the full Stelle's gravity to various orders or variations of $R+R^2$ theories. It is worth mentioning that, as said before, theories with odd powers of $c$ will be equivalent to one of the theories above, and higher powers of $c$ may be problematic in the Galilean limit. Note that the LO actions possess Carrollian symmetries by construction so they are Carrollian theories, but the NLO action do not. The NLO of the Carrollian expansion does not preserve Carrollian symmetry in general, however, certain truncation recovers the symmetries resulting in the magnetic Carrollian limit of the theory.}
\label{table:2}
\end{table}

Focusing on the ghost-free theories, namely (2,4) and (4,2), we see that (2,4) is the same as GR to the LO, so the electric limit and the constraints to the magnetic limit are the same as those of GR. However, to the NLO the theory has extra terms which can be interpreted as an additional flux. In the case of (4,2) the LO and the NLO are equivalent to that of $R+\beta'R^2$ theory. The constraints and the evolution equations are in general much more complicated. However, there is a special case where the LO equations reduce to GR with a cosmological constant, this means that the full theory gives rise to an emergent cosmological constant in addition to the extra terms which, like in the (2,4) case, can be interpreted as an additional flux.

More work has to be done to study the field equations for these theories to the LO and NLO. It would be interesting to compare each case with GR to understand what modifications can arise from different quartic terms of the extrinsic curvature. Another direction for future research is to calculate the Galilean limit of quadratic gravity. Since the dependence of $\alpha$ and $\beta$ on $c$ is not a perturbative assumption, the higher powers of $c$ in the action may be problematic in the Galilean limit. In the current classification the most attractive options for future study are $(2,4)$ and $(4,2)$ since, after imposing the tachyon removing conditions, we get the Carrollian limit of $R+\beta R^2$, a renormalizable theory with no ghosts or tachyons (only if $\beta$ is positive) which is deduced directly from the string theory. We plan to study black hole solutions for these theories. Since $R+R^2$ theories have more black hole solutions than GR, a direction for a future work is to study black hole solutions for their actions. These should coincide with the Carrollian limit of Schwarzchild-Bach solutions. it is interesting to analyze the dynamics of Carrollian particles on horizons of various black-hole solutions and compare the dynamics with that of \cite{Bicak:2023vxs} and study the modifications arising from the quartic terms. 

%%%%%%%%%%%%%%%%%%%%%%%%%%%%%%%%%%%%%%%%%%%%%%%%%%%%%%%%%%%%%%%%%%%%%%%%%%%%%%%%%%%%%%
%% ACKNOWLEDGEMENTS

\section*{Acknowledgements}

The authors would like to thank Eric Bergshoeff (Groningen, Netherlands), Pavel Krtou\v{s}, David Kubiz\v{n}\'ak (Prague, Czechia), and Marc Henneaux (Brussels, Belgium) for stimulating discussions. P.T. and I.K. were supported by Primus grant PRIMUS/23/SCI/005 from Charles University.

%%%%%%%%%%%%%%%%%%%%%%%%%%%%%%%%%%%%%%%%%%%%%%%%%%%%%%%%%%%%%%%%%%%%%%%%%%%%%%%%%%%%%%
%% Appendix

\appendix


\begin{thebibliography}{82}%
\makeatletter
\providecommand \@ifxundefined [1]{%
 \@ifx{#1\undefined}
}%
\providecommand \@ifnum [1]{%
 \ifnum #1\expandafter \@firstoftwo
 \else \expandafter \@secondoftwo
 \fi
}%
\providecommand \@ifx [1]{%
 \ifx #1\expandafter \@firstoftwo
 \else \expandafter \@secondoftwo
 \fi
}%
\providecommand \natexlab [1]{#1}%
\providecommand \enquote  [1]{``#1''}%
\providecommand \bibnamefont  [1]{#1}%
\providecommand \bibfnamefont [1]{#1}%
\providecommand \citenamefont [1]{#1}%
\providecommand \href@noop [0]{\@secondoftwo}%
\providecommand \href [0]{\begingroup \@sanitize@url \@href}%
\providecommand \@href[1]{\@@startlink{#1}\@@href}%
\providecommand \@@href[1]{\endgroup#1\@@endlink}%
\providecommand \@sanitize@url [0]{\catcode `\\12\catcode `\$12\catcode
  `\&12\catcode `\#12\catcode `\^12\catcode `\_12\catcode `\%12\relax}%
\providecommand \@@startlink[1]{}%
\providecommand \@@endlink[0]{}%
\providecommand \url  [0]{\begingroup\@sanitize@url \@url }%
\providecommand \@url [1]{\endgroup\@href {#1}{\urlprefix }}%
\providecommand \urlprefix  [0]{URL }%
\providecommand \Eprint [0]{\href }%
\providecommand \doibase [0]{https://doi.org/}%
\providecommand \selectlanguage [0]{\@gobble}%
\providecommand \bibinfo  [0]{\@secondoftwo}%
\providecommand \bibfield  [0]{\@secondoftwo}%
\providecommand \translation [1]{[#1]}%
\providecommand \BibitemOpen [0]{}%
\providecommand \bibitemStop [0]{}%
\providecommand \bibitemNoStop [0]{.\EOS\space}%
\providecommand \EOS [0]{\spacefactor3000\relax}%
\providecommand \BibitemShut  [1]{\csname bibitem#1\endcsname}%
\let\auto@bib@innerbib\@empty
%</preamble>
\bibitem [{\citenamefont {Zumino}(1986)}]{ZUMINO1986109}%
  \BibitemOpen
  \bibfield  {author} {\bibinfo {author} {\bibfnamefont {B.}~\bibnamefont
  {Zumino}},\ }\bibfield  {title} {\bibinfo {title} {Gravity theories in more
  than four dimensions},\ }\href
  {https://doi.org/https://doi.org/10.1016/0370-1573(86)90076-1} {\bibfield
  {journal} {\bibinfo  {journal} {Physics Reports}\ }\textbf {\bibinfo {volume}
  {137}},\ \bibinfo {pages} {109} (\bibinfo {year} {1986})}\BibitemShut
  {NoStop}%
\bibitem [{\citenamefont {Zwiebach}(1985)}]{ZWIEBACH1985315}%
  \BibitemOpen
  \bibfield  {author} {\bibinfo {author} {\bibfnamefont {B.}~\bibnamefont
  {Zwiebach}},\ }\bibfield  {title} {\bibinfo {title} {Curvature squared terms
  and string theories},\ }\href
  {https://doi.org/https://doi.org/10.1016/0370-2693(85)91616-8} {\bibfield
  {journal} {\bibinfo  {journal} {Physics Letters B}\ }\textbf {\bibinfo
  {volume} {156}},\ \bibinfo {pages} {315} (\bibinfo {year}
  {1985})}\BibitemShut {NoStop}%
\bibitem [{\citenamefont {Forger}\ \emph {et~al.}(1996)\citenamefont {Forger},
  \citenamefont {Ovrut}, \citenamefont {Theisen},\ and\ \citenamefont
  {Waldram}}]{Forger:1996vj}%
  \BibitemOpen
  \bibfield  {author} {\bibinfo {author} {\bibfnamefont {K.}~\bibnamefont
  {Forger}}, \bibinfo {author} {\bibfnamefont {B.~A.}\ \bibnamefont {Ovrut}},
  \bibinfo {author} {\bibfnamefont {S.~J.}\ \bibnamefont {Theisen}},\ and\
  \bibinfo {author} {\bibfnamefont {D.}~\bibnamefont {Waldram}},\ }\bibfield
  {title} {\bibinfo {title} {{Higher derivative gravity in string theory}},\
  }\href {https://doi.org/10.1016/S0370-2693(96)01175-6} {\bibfield  {journal}
  {\bibinfo  {journal} {Phys. Lett. B}\ }\textbf {\bibinfo {volume} {388}},\
  \bibinfo {pages} {512} (\bibinfo {year} {1996})},\ \Eprint
  {https://arxiv.org/abs/hep-th/9605145} {arXiv:hep-th/9605145} \BibitemShut
  {NoStop}%
\bibitem [{\citenamefont {Myers}(1987)}]{Myers:1987yn}%
  \BibitemOpen
  \bibfield  {author} {\bibinfo {author} {\bibfnamefont {R.~C.}\ \bibnamefont
  {Myers}},\ }\bibfield  {title} {\bibinfo {title} {{Higher Derivative Gravity,
  Surface Terms and String Theory}},\ }\href
  {https://doi.org/10.1103/PhysRevD.36.392} {\bibfield  {journal} {\bibinfo
  {journal} {Phys. Rev. D}\ }\textbf {\bibinfo {volume} {36}},\ \bibinfo
  {pages} {392} (\bibinfo {year} {1987})}\BibitemShut {NoStop}%
\bibitem [{\citenamefont {Alvarez-Gaume}\ \emph {et~al.}(2016)\citenamefont
  {Alvarez-Gaume}, \citenamefont {Kehagias}, \citenamefont {Kounnas},
  \citenamefont {L\"ust},\ and\ \citenamefont
  {Riotto}}]{Alvarez-Gaume:2015rwa}%
  \BibitemOpen
  \bibfield  {author} {\bibinfo {author} {\bibfnamefont {L.}~\bibnamefont
  {Alvarez-Gaume}}, \bibinfo {author} {\bibfnamefont {A.}~\bibnamefont
  {Kehagias}}, \bibinfo {author} {\bibfnamefont {C.}~\bibnamefont {Kounnas}},
  \bibinfo {author} {\bibfnamefont {D.}~\bibnamefont {L\"ust}},\ and\ \bibinfo
  {author} {\bibfnamefont {A.}~\bibnamefont {Riotto}},\ }\bibfield  {title}
  {\bibinfo {title} {{Aspects of Quadratic Gravity}},\ }\href
  {https://doi.org/10.1002/prop.201500100} {\bibfield  {journal} {\bibinfo
  {journal} {Fortsch. Phys.}\ }\textbf {\bibinfo {volume} {64}},\ \bibinfo
  {pages} {176} (\bibinfo {year} {2016})},\ \Eprint
  {https://arxiv.org/abs/1505.07657} {arXiv:1505.07657 [hep-th]} \BibitemShut
  {NoStop}%
\bibitem [{\citenamefont {Nenmeli}\ \emph {et~al.}(2021)\citenamefont
  {Nenmeli}, \citenamefont {Shankaranarayanan}, \citenamefont {Todorinov},\
  and\ \citenamefont {Das}}]{Nenmeli:2021orl}%
  \BibitemOpen
  \bibfield  {author} {\bibinfo {author} {\bibfnamefont {V.}~\bibnamefont
  {Nenmeli}}, \bibinfo {author} {\bibfnamefont {S.}~\bibnamefont
  {Shankaranarayanan}}, \bibinfo {author} {\bibfnamefont {V.}~\bibnamefont
  {Todorinov}},\ and\ \bibinfo {author} {\bibfnamefont {S.}~\bibnamefont
  {Das}},\ }\bibfield  {title} {\bibinfo {title} {{Maximal momentum GUP leads
  to quadratic gravity}},\ }\href
  {https://doi.org/10.1016/j.physletb.2021.136621} {\bibfield  {journal}
  {\bibinfo  {journal} {Phys. Lett. B}\ }\textbf {\bibinfo {volume} {821}},\
  \bibinfo {pages} {136621} (\bibinfo {year} {2021})},\ \Eprint
  {https://arxiv.org/abs/2106.04141} {arXiv:2106.04141 [gr-qc]} \BibitemShut
  {NoStop}%
\bibitem [{\citenamefont {Stelle}(1978)}]{Stelle:1977ry}%
  \BibitemOpen
  \bibfield  {author} {\bibinfo {author} {\bibfnamefont {K.~S.}\ \bibnamefont
  {Stelle}},\ }\bibfield  {title} {\bibinfo {title} {{Classical Gravity with
  Higher Derivatives}},\ }\href {https://doi.org/10.1007/BF00760427} {\bibfield
   {journal} {\bibinfo  {journal} {Gen. Rel. Grav.}\ }\textbf {\bibinfo
  {volume} {9}},\ \bibinfo {pages} {353} (\bibinfo {year} {1978})}\BibitemShut
  {NoStop}%
\bibitem [{\citenamefont {Stelle}(1977)}]{Stelle:1976gc}%
  \BibitemOpen
  \bibfield  {author} {\bibinfo {author} {\bibfnamefont {K.~S.}\ \bibnamefont
  {Stelle}},\ }\bibfield  {title} {\bibinfo {title} {{Renormalization of Higher
  Derivative Quantum Gravity}},\ }\href
  {https://doi.org/10.1103/PhysRevD.16.953} {\bibfield  {journal} {\bibinfo
  {journal} {Phys. Rev. D}\ }\textbf {\bibinfo {volume} {16}},\ \bibinfo
  {pages} {953} (\bibinfo {year} {1977})}\BibitemShut {NoStop}%
\bibitem [{\citenamefont {Julve}\ and\ \citenamefont
  {Tonin}(1978)}]{Julve:1978xn}%
  \BibitemOpen
  \bibfield  {author} {\bibinfo {author} {\bibfnamefont {J.}~\bibnamefont
  {Julve}}\ and\ \bibinfo {author} {\bibfnamefont {M.}~\bibnamefont {Tonin}},\
  }\bibfield  {title} {\bibinfo {title} {{Quantum Gravity with Higher
  Derivative Terms}},\ }\href {https://doi.org/10.1007/BF02748637} {\bibfield
  {journal} {\bibinfo  {journal} {Nuovo Cim. B}\ }\textbf {\bibinfo {volume}
  {46}},\ \bibinfo {pages} {137} (\bibinfo {year} {1978})}\BibitemShut
  {NoStop}%
\bibitem [{\citenamefont {Lu}\ \emph {et~al.}(2015)\citenamefont {Lu},
  \citenamefont {Perkins}, \citenamefont {Pope},\ and\ \citenamefont
  {Stelle}}]{Lu:2015cqa}%
  \BibitemOpen
  \bibfield  {author} {\bibinfo {author} {\bibfnamefont {H.}~\bibnamefont
  {Lu}}, \bibinfo {author} {\bibfnamefont {A.}~\bibnamefont {Perkins}},
  \bibinfo {author} {\bibfnamefont {C.~N.}\ \bibnamefont {Pope}},\ and\
  \bibinfo {author} {\bibfnamefont {K.~S.}\ \bibnamefont {Stelle}},\ }\bibfield
   {title} {\bibinfo {title} {{Black Holes in Higher-Derivative Gravity}},\
  }\href {https://doi.org/10.1103/PhysRevLett.114.171601} {\bibfield  {journal}
  {\bibinfo  {journal} {Phys. Rev. Lett.}\ }\textbf {\bibinfo {volume} {114}},\
  \bibinfo {pages} {171601} (\bibinfo {year} {2015})},\ \Eprint
  {https://arxiv.org/abs/1502.01028} {arXiv:1502.01028 [hep-th]} \BibitemShut
  {NoStop}%
\bibitem [{\citenamefont {L\"u}\ \emph {et~al.}(2015)\citenamefont {L\"u},
  \citenamefont {Perkins}, \citenamefont {Pope},\ and\ \citenamefont
  {Stelle}}]{Lu:2015psa}%
  \BibitemOpen
  \bibfield  {author} {\bibinfo {author} {\bibfnamefont {H.}~\bibnamefont
  {L\"u}}, \bibinfo {author} {\bibfnamefont {A.}~\bibnamefont {Perkins}},
  \bibinfo {author} {\bibfnamefont {C.~N.}\ \bibnamefont {Pope}},\ and\
  \bibinfo {author} {\bibfnamefont {K.~S.}\ \bibnamefont {Stelle}},\ }\bibfield
   {title} {\bibinfo {title} {{Spherically Symmetric Solutions in
  Higher-Derivative Gravity}},\ }\href
  {https://doi.org/10.1103/PhysRevD.92.124019} {\bibfield  {journal} {\bibinfo
  {journal} {Phys. Rev. D}\ }\textbf {\bibinfo {volume} {92}},\ \bibinfo
  {pages} {124019} (\bibinfo {year} {2015})},\ \Eprint
  {https://arxiv.org/abs/1508.00010} {arXiv:1508.00010 [hep-th]} \BibitemShut
  {NoStop}%
\bibitem [{\citenamefont {Podolsk\'y}\ \emph {et~al.}(2020)\citenamefont
  {Podolsk\'y}, \citenamefont {\v{S}varc}, \citenamefont {Pravda},\ and\
  \citenamefont {Pravdova}}]{Podolsky:2019gro}%
  \BibitemOpen
  \bibfield  {author} {\bibinfo {author} {\bibfnamefont {J.}~\bibnamefont
  {Podolsk\'y}}, \bibinfo {author} {\bibfnamefont {R.}~\bibnamefont
  {\v{S}varc}}, \bibinfo {author} {\bibfnamefont {V.}~\bibnamefont {Pravda}},\
  and\ \bibinfo {author} {\bibfnamefont {A.}~\bibnamefont {Pravdova}},\
  }\bibfield  {title} {\bibinfo {title} {{Black holes and other exact spherical
  solutions in Quadratic Gravity}},\ }\href
  {https://doi.org/10.1103/PhysRevD.101.024027} {\bibfield  {journal} {\bibinfo
   {journal} {Phys. Rev. D}\ }\textbf {\bibinfo {volume} {101}},\ \bibinfo
  {pages} {024027} (\bibinfo {year} {2020})},\ \Eprint
  {https://arxiv.org/abs/1907.00046} {arXiv:1907.00046 [gr-qc]} \BibitemShut
  {NoStop}%
\bibitem [{\citenamefont {Pravdova}\ \emph {et~al.}(2023)\citenamefont
  {Pravdova}, \citenamefont {Pravda},\ and\ \citenamefont
  {Ortaggio}}]{Pravdova:2023nbo}%
  \BibitemOpen
  \bibfield  {author} {\bibinfo {author} {\bibfnamefont {A.}~\bibnamefont
  {Pravdova}}, \bibinfo {author} {\bibfnamefont {V.}~\bibnamefont {Pravda}},\
  and\ \bibinfo {author} {\bibfnamefont {M.}~\bibnamefont {Ortaggio}},\
  }\bibfield  {title} {\bibinfo {title} {{Topological black holes in higher
  derivative gravity}},\ }\href
  {https://doi.org/10.1140/epjc/s10052-023-11338-9} {\bibfield  {journal}
  {\bibinfo  {journal} {Eur. Phys. J. C}\ }\textbf {\bibinfo {volume} {83}},\
  \bibinfo {pages} {180} (\bibinfo {year} {2023})},\ \Eprint
  {https://arxiv.org/abs/2301.10720} {arXiv:2301.10720 [gr-qc]} \BibitemShut
  {NoStop}%
\bibitem [{\citenamefont {Levy-Leblond}(1965)}]{Levy-Leblond}%
  \BibitemOpen
  \bibfield  {author} {\bibinfo {author} {\bibfnamefont {J.-M.}\ \bibnamefont
  {Levy-Leblond}},\ }\bibfield  {title} {\bibinfo {title} {{Une nouvelle limite
  non-relativiste du groupe de Poincaré}},\ }\href
  {https://doi.org/http://archive.numdam.org/item/AIHPA_1965__3_1_1_0/}
  {\bibfield  {journal} {\bibinfo  {journal} {Annales de l'institut Henri
  Poincaré. Section A, Physique Théorique}\ }\textbf {\bibinfo {volume}
  {3}},\ \bibinfo {pages} {1} (\bibinfo {year} {1965})}\BibitemShut {NoStop}%
\bibitem [{\citenamefont {Sen~Gupta}(1966)}]{Gupta}%
  \BibitemOpen
  \bibfield  {author} {\bibinfo {author} {\bibfnamefont {N.}~\bibnamefont
  {Sen~Gupta}},\ }\bibfield  {title} {\bibinfo {title} {On an analogue of the
  galilei group},\ }\href {https://doi.org/https://doi.org/10.1007/BF02740871}
  {\bibfield  {journal} {\bibinfo  {journal} {Nuovo Cimento A (1965-1970)}\
  }\textbf {\bibinfo {volume} {44}},\ \bibinfo {pages} {512} (\bibinfo {year}
  {1966})}\BibitemShut {NoStop}%
\bibitem [{\citenamefont {Zhang}\ \emph {et~al.}(2023)\citenamefont {Zhang},
  \citenamefont {Zeng},\ and\ \citenamefont {Horvathy}}]{Zhang:2023jbi}%
  \BibitemOpen
  \bibfield  {author} {\bibinfo {author} {\bibfnamefont {P.~M.}\ \bibnamefont
  {Zhang}}, \bibinfo {author} {\bibfnamefont {H.-X.}\ \bibnamefont {Zeng}},\
  and\ \bibinfo {author} {\bibfnamefont {P.~A.}\ \bibnamefont {Horvathy}},\
  }\href@noop {} {\bibinfo {title} {{MultiCarroll dynamics}}} (\bibinfo {year}
  {2023}),\ \Eprint {https://arxiv.org/abs/2306.07002} {arXiv:2306.07002
  [gr-qc]} \BibitemShut {NoStop}%
\bibitem [{\citenamefont {Bergshoeff}\ \emph {et~al.}(2014)\citenamefont
  {Bergshoeff}, \citenamefont {Gomis},\ and\ \citenamefont
  {Longhi}}]{Bergshoeff_2014}%
  \BibitemOpen
  \bibfield  {author} {\bibinfo {author} {\bibfnamefont {E.}~\bibnamefont
  {Bergshoeff}}, \bibinfo {author} {\bibfnamefont {J.}~\bibnamefont {Gomis}},\
  and\ \bibinfo {author} {\bibfnamefont {G.}~\bibnamefont {Longhi}},\
  }\bibfield  {title} {\bibinfo {title} {Dynamics of carroll particles},\
  }\href {https://doi.org/10.1088/0264-9381/31/20/205009} {\bibfield  {journal}
  {\bibinfo  {journal} {Classical and Quantum Gravity}\ }\textbf {\bibinfo
  {volume} {31}},\ \bibinfo {pages} {205009} (\bibinfo {year}
  {2014})}\BibitemShut {NoStop}%
\bibitem [{\citenamefont {Marsot}(2022)}]{Marsot:2021tvq}%
  \BibitemOpen
  \bibfield  {author} {\bibinfo {author} {\bibfnamefont {L.}~\bibnamefont
  {Marsot}},\ }\bibfield  {title} {\bibinfo {title} {{Planar Carrollean
  dynamics, and the Carroll quantum equation}},\ }\href
  {https://doi.org/10.1016/j.geomphys.2022.104574} {\bibfield  {journal}
  {\bibinfo  {journal} {J. Geom. Phys.}\ }\textbf {\bibinfo {volume} {179}},\
  \bibinfo {pages} {104574} (\bibinfo {year} {2022})},\ \Eprint
  {https://arxiv.org/abs/2110.08489} {arXiv:2110.08489 [math-ph]} \BibitemShut
  {NoStop}%
\bibitem [{\citenamefont {Marsot}\ \emph
  {et~al.}(2022{\natexlab{a}})\citenamefont {Marsot}, \citenamefont {Zhang},
  \citenamefont {Chernodub},\ and\ \citenamefont {Horvathy}}]{Marsot:2022imf}%
  \BibitemOpen
  \bibfield  {author} {\bibinfo {author} {\bibfnamefont {L.}~\bibnamefont
  {Marsot}}, \bibinfo {author} {\bibfnamefont {P.~M.}\ \bibnamefont {Zhang}},
  \bibinfo {author} {\bibfnamefont {M.}~\bibnamefont {Chernodub}},\ and\
  \bibinfo {author} {\bibfnamefont {P.~A.}\ \bibnamefont {Horvathy}},\
  }\href@noop {} {\bibinfo {title} {{Hall effects in Carroll dynamics}}}
  (\bibinfo {year} {2022}{\natexlab{a}}),\ \Eprint
  {https://arxiv.org/abs/2212.02360} {arXiv:2212.02360 [hep-th]} \BibitemShut
  {NoStop}%
\bibitem [{\citenamefont {Bagchi}\ \emph
  {et~al.}(2023{\natexlab{a}})\citenamefont {Bagchi}, \citenamefont {Banerjee},
  \citenamefont {Basu}, \citenamefont {Islam},\ and\ \citenamefont
  {Mondal}}]{Bagchi:2022eui}%
  \BibitemOpen
  \bibfield  {author} {\bibinfo {author} {\bibfnamefont {A.}~\bibnamefont
  {Bagchi}}, \bibinfo {author} {\bibfnamefont {A.}~\bibnamefont {Banerjee}},
  \bibinfo {author} {\bibfnamefont {R.}~\bibnamefont {Basu}}, \bibinfo {author}
  {\bibfnamefont {M.}~\bibnamefont {Islam}},\ and\ \bibinfo {author}
  {\bibfnamefont {S.}~\bibnamefont {Mondal}},\ }\bibfield  {title} {\bibinfo
  {title} {{Magic fermions: Carroll and flat bands}},\ }\href
  {https://doi.org/10.1007/JHEP03(2023)227} {\bibfield  {journal} {\bibinfo
  {journal} {JHEP}\ }\textbf {\bibinfo {volume} {03}},\ \bibinfo {pages}
  {227}},\ \Eprint {https://arxiv.org/abs/2211.11640} {arXiv:2211.11640
  [hep-th]} \BibitemShut {NoStop}%
\bibitem [{\citenamefont {Kubakaddi}(2021)}]{Kubakaddi_2021}%
  \BibitemOpen
  \bibfield  {author} {\bibinfo {author} {\bibfnamefont {S.~S.}\ \bibnamefont
  {Kubakaddi}},\ }\bibfield  {title} {\bibinfo {title} {Giant thermopower and
  power factor in magic angle twisted bilayer graphene at low temperature},\
  }\href {https://doi.org/10.1088/1361-648x/abf0c2} {\bibfield  {journal}
  {\bibinfo  {journal} {Journal of Physics: Condensed Matter}\ }\textbf
  {\bibinfo {volume} {33}},\ \bibinfo {pages} {245704} (\bibinfo {year}
  {2021})}\BibitemShut {NoStop}%
\bibitem [{\citenamefont {Kononov}\ \emph {et~al.}(2021)\citenamefont
  {Kononov}, \citenamefont {Endres}, \citenamefont {Abulizi}, \citenamefont
  {Qu}, \citenamefont {Yan}, \citenamefont {Mandrus}, \citenamefont {Watanabe},
  \citenamefont {Taniguchi},\ and\ \citenamefont
  {Schönenberger}}]{Kononov_2021}%
  \BibitemOpen
  \bibfield  {author} {\bibinfo {author} {\bibfnamefont {A.}~\bibnamefont
  {Kononov}}, \bibinfo {author} {\bibfnamefont {M.}~\bibnamefont {Endres}},
  \bibinfo {author} {\bibfnamefont {G.}~\bibnamefont {Abulizi}}, \bibinfo
  {author} {\bibfnamefont {K.}~\bibnamefont {Qu}}, \bibinfo {author}
  {\bibfnamefont {J.}~\bibnamefont {Yan}}, \bibinfo {author} {\bibfnamefont
  {D.~G.}\ \bibnamefont {Mandrus}}, \bibinfo {author} {\bibfnamefont
  {K.}~\bibnamefont {Watanabe}}, \bibinfo {author} {\bibfnamefont
  {T.}~\bibnamefont {Taniguchi}},\ and\ \bibinfo {author} {\bibfnamefont
  {C.}~\bibnamefont {Schönenberger}},\ }\bibfield  {title} {\bibinfo {title}
  {Superconductivity in type-{II} weyl-semimetal \text{WTe}$_2$ induced by a
  normal metal contact},\ }\href {https://doi.org/10.1063/5.0021350} {\bibfield
   {journal} {\bibinfo  {journal} {Journal of Applied Physics}\ }\textbf
  {\bibinfo {volume} {129}},\ \bibinfo {pages} {113903} (\bibinfo {year}
  {2021})}\BibitemShut {NoStop}%
\bibitem [{\citenamefont {Rivera-Betancour}\ and\ \citenamefont
  {Vilatte}(2022)}]{PhysRevD.106.085004}%
  \BibitemOpen
  \bibfield  {author} {\bibinfo {author} {\bibfnamefont {D.}~\bibnamefont
  {Rivera-Betancour}}\ and\ \bibinfo {author} {\bibfnamefont {M.}~\bibnamefont
  {Vilatte}},\ }\bibfield  {title} {\bibinfo {title} {Revisiting the carrollian
  scalar field},\ }\href {https://doi.org/10.1103/PhysRevD.106.085004}
  {\bibfield  {journal} {\bibinfo  {journal} {Phys. Rev. D}\ }\textbf {\bibinfo
  {volume} {106}},\ \bibinfo {pages} {085004} (\bibinfo {year}
  {2022})}\BibitemShut {NoStop}%
\bibitem [{\citenamefont {Chen}\ \emph {et~al.}(2023)\citenamefont {Chen},
  \citenamefont {Liu}, \citenamefont {Sun},\ and\ \citenamefont
  {Zheng}}]{Chen:2023pqf}%
  \BibitemOpen
  \bibfield  {author} {\bibinfo {author} {\bibfnamefont {B.}~\bibnamefont
  {Chen}}, \bibinfo {author} {\bibfnamefont {R.}~\bibnamefont {Liu}}, \bibinfo
  {author} {\bibfnamefont {H.}~\bibnamefont {Sun}},\ and\ \bibinfo {author}
  {\bibfnamefont {Y.-f.}\ \bibnamefont {Zheng}},\ }\href@noop {} {\bibinfo
  {title} {{Constructing Carrollian Field Theories from Null Reduction}}}
  (\bibinfo {year} {2023}),\ \Eprint {https://arxiv.org/abs/2301.06011}
  {arXiv:2301.06011 [hep-th]} \BibitemShut {NoStop}%
\bibitem [{\citenamefont {Bergshoeff}\ \emph {et~al.}(2022)\citenamefont
  {Bergshoeff}, \citenamefont {Figueroa-O'Farrill},\ and\ \citenamefont
  {Gomis}}]{Bergshoeff:2022eog}%
  \BibitemOpen
  \bibfield  {author} {\bibinfo {author} {\bibfnamefont {E.}~\bibnamefont
  {Bergshoeff}}, \bibinfo {author} {\bibfnamefont {J.}~\bibnamefont
  {Figueroa-O'Farrill}},\ and\ \bibinfo {author} {\bibfnamefont
  {J.}~\bibnamefont {Gomis}},\ }\href@noop {} {\bibinfo {title} {{A
  non-lorentzian primer}}} (\bibinfo {year} {2022}),\ \Eprint
  {https://arxiv.org/abs/2206.12177} {arXiv:2206.12177 [hep-th]} \BibitemShut
  {NoStop}%
\bibitem [{\citenamefont {Henneaux}\ and\ \citenamefont
  {Salgado-Rebolledo}(2021)}]{Henneaux:2021yzg}%
  \BibitemOpen
  \bibfield  {author} {\bibinfo {author} {\bibfnamefont {M.}~\bibnamefont
  {Henneaux}}\ and\ \bibinfo {author} {\bibfnamefont {P.}~\bibnamefont
  {Salgado-Rebolledo}},\ }\bibfield  {title} {\bibinfo {title} {{Carroll
  contractions of Lorentz-invariant theories}},\ }\href
  {https://doi.org/10.1007/JHEP11(2021)180} {\bibfield  {journal} {\bibinfo
  {journal} {JHEP}\ }\textbf {\bibinfo {volume} {11}},\ \bibinfo {pages}
  {180}},\ \Eprint {https://arxiv.org/abs/2109.06708} {arXiv:2109.06708
  [hep-th]} \BibitemShut {NoStop}%
\bibitem [{\citenamefont {Bagchi}\ \emph
  {et~al.}(2019{\natexlab{a}})\citenamefont {Bagchi}, \citenamefont {Mehra},\
  and\ \citenamefont {Nandi}}]{Bagchi:2019xfx}%
  \BibitemOpen
  \bibfield  {author} {\bibinfo {author} {\bibfnamefont {A.}~\bibnamefont
  {Bagchi}}, \bibinfo {author} {\bibfnamefont {A.}~\bibnamefont {Mehra}},\ and\
  \bibinfo {author} {\bibfnamefont {P.}~\bibnamefont {Nandi}},\ }\bibfield
  {title} {\bibinfo {title} {{Field Theories with Conformal Carrollian
  Symmetry}},\ }\href {https://doi.org/10.1007/JHEP05(2019)108} {\bibfield
  {journal} {\bibinfo  {journal} {JHEP}\ }\textbf {\bibinfo {volume} {05}},\
  \bibinfo {pages} {108}},\ \Eprint {https://arxiv.org/abs/1901.10147}
  {arXiv:1901.10147 [hep-th]} \BibitemShut {NoStop}%
\bibitem [{\citenamefont {Bagchi}\ \emph {et~al.}(2020)\citenamefont {Bagchi},
  \citenamefont {Basu}, \citenamefont {Mehra},\ and\ \citenamefont
  {Nandi}}]{Bagchi:2019clu}%
  \BibitemOpen
  \bibfield  {author} {\bibinfo {author} {\bibfnamefont {A.}~\bibnamefont
  {Bagchi}}, \bibinfo {author} {\bibfnamefont {R.}~\bibnamefont {Basu}},
  \bibinfo {author} {\bibfnamefont {A.}~\bibnamefont {Mehra}},\ and\ \bibinfo
  {author} {\bibfnamefont {P.}~\bibnamefont {Nandi}},\ }\bibfield  {title}
  {\bibinfo {title} {{Field Theories on Null Manifolds}},\ }\href
  {https://doi.org/10.1007/JHEP02(2020)141} {\bibfield  {journal} {\bibinfo
  {journal} {JHEP}\ }\textbf {\bibinfo {volume} {02}},\ \bibinfo {pages}
  {141}},\ \Eprint {https://arxiv.org/abs/1912.09388} {arXiv:1912.09388
  [hep-th]} \BibitemShut {NoStop}%
\bibitem [{\citenamefont {Bagchi}\ \emph {et~al.}(2021)\citenamefont {Bagchi},
  \citenamefont {Dutta}, \citenamefont {Kolekar},\ and\ \citenamefont
  {Sharma}}]{Bagchi:2021gai}%
  \BibitemOpen
  \bibfield  {author} {\bibinfo {author} {\bibfnamefont {A.}~\bibnamefont
  {Bagchi}}, \bibinfo {author} {\bibfnamefont {S.}~\bibnamefont {Dutta}},
  \bibinfo {author} {\bibfnamefont {K.~S.}\ \bibnamefont {Kolekar}},\ and\
  \bibinfo {author} {\bibfnamefont {P.}~\bibnamefont {Sharma}},\ }\bibfield
  {title} {\bibinfo {title} {{BMS field theories and Weyl anomaly}},\ }\href
  {https://doi.org/10.1007/JHEP07(2021)101} {\bibfield  {journal} {\bibinfo
  {journal} {JHEP}\ }\textbf {\bibinfo {volume} {07}},\ \bibinfo {pages}
  {101}},\ \Eprint {https://arxiv.org/abs/2104.10405} {arXiv:2104.10405
  [hep-th]} \BibitemShut {NoStop}%
\bibitem [{\citenamefont {Banerjee}\ \emph {et~al.}(2021)\citenamefont
  {Banerjee}, \citenamefont {Basu}, \citenamefont {Mehra}, \citenamefont
  {Mohan},\ and\ \citenamefont {Sharma}}]{PhysRevD.103.105001}%
  \BibitemOpen
  \bibfield  {author} {\bibinfo {author} {\bibfnamefont {K.}~\bibnamefont
  {Banerjee}}, \bibinfo {author} {\bibfnamefont {R.}~\bibnamefont {Basu}},
  \bibinfo {author} {\bibfnamefont {A.}~\bibnamefont {Mehra}}, \bibinfo
  {author} {\bibfnamefont {A.}~\bibnamefont {Mohan}},\ and\ \bibinfo {author}
  {\bibfnamefont {A.}~\bibnamefont {Sharma}},\ }\bibfield  {title} {\bibinfo
  {title} {Interacting conformal carrollian theories: Cues from
  electrodynamics},\ }\href {https://doi.org/10.1103/PhysRevD.103.105001}
  {\bibfield  {journal} {\bibinfo  {journal} {Phys. Rev. D}\ }\textbf {\bibinfo
  {volume} {103}},\ \bibinfo {pages} {105001} (\bibinfo {year}
  {2021})}\BibitemShut {NoStop}%
\bibitem [{\citenamefont {Bagchi}\ \emph
  {et~al.}(2023{\natexlab{b}})\citenamefont {Bagchi}, \citenamefont {Kolekar},\
  and\ \citenamefont {Shukla}}]{Bagchi:2023ysc}%
  \BibitemOpen
  \bibfield  {author} {\bibinfo {author} {\bibfnamefont {A.}~\bibnamefont
  {Bagchi}}, \bibinfo {author} {\bibfnamefont {K.~S.}\ \bibnamefont
  {Kolekar}},\ and\ \bibinfo {author} {\bibfnamefont {A.}~\bibnamefont
  {Shukla}},\ }\bibfield  {title} {\bibinfo {title} {{Carrollian Origins of
  Bjorken Flow}},\ }\href {https://doi.org/10.1103/PhysRevLett.130.241601}
  {\bibfield  {journal} {\bibinfo  {journal} {Phys. Rev. Lett.}\ }\textbf
  {\bibinfo {volume} {130}},\ \bibinfo {pages} {241601} (\bibinfo {year}
  {2023}{\natexlab{b}})},\ \Eprint {https://arxiv.org/abs/2302.03053}
  {arXiv:2302.03053 [hep-th]} \BibitemShut {NoStop}%
\bibitem [{\citenamefont {Ciambelli}\ \emph
  {et~al.}(2018{\natexlab{a}})\citenamefont {Ciambelli}, \citenamefont
  {Marteau}, \citenamefont {Petkou}, \citenamefont {Petropoulos},\ and\
  \citenamefont {Siampos}}]{Ciambelli:2018wre}%
  \BibitemOpen
  \bibfield  {author} {\bibinfo {author} {\bibfnamefont {L.}~\bibnamefont
  {Ciambelli}}, \bibinfo {author} {\bibfnamefont {C.}~\bibnamefont {Marteau}},
  \bibinfo {author} {\bibfnamefont {A.~C.}\ \bibnamefont {Petkou}}, \bibinfo
  {author} {\bibfnamefont {P.~M.}\ \bibnamefont {Petropoulos}},\ and\ \bibinfo
  {author} {\bibfnamefont {K.}~\bibnamefont {Siampos}},\ }\bibfield  {title}
  {\bibinfo {title} {{Flat holography and Carrollian fluids}},\ }\href
  {https://doi.org/10.1007/JHEP07(2018)165} {\bibfield  {journal} {\bibinfo
  {journal} {JHEP}\ }\textbf {\bibinfo {volume} {07}},\ \bibinfo {pages}
  {165}},\ \Eprint {https://arxiv.org/abs/1802.06809} {arXiv:1802.06809
  [hep-th]} \BibitemShut {NoStop}%
\bibitem [{\citenamefont {Ciambelli}\ \emph
  {et~al.}(2018{\natexlab{b}})\citenamefont {Ciambelli}, \citenamefont
  {Marteau}, \citenamefont {Petkou}, \citenamefont {Petropoulos},\ and\
  \citenamefont {Siampos}}]{Ciambelli:2018xat}%
  \BibitemOpen
  \bibfield  {author} {\bibinfo {author} {\bibfnamefont {L.}~\bibnamefont
  {Ciambelli}}, \bibinfo {author} {\bibfnamefont {C.}~\bibnamefont {Marteau}},
  \bibinfo {author} {\bibfnamefont {A.~C.}\ \bibnamefont {Petkou}}, \bibinfo
  {author} {\bibfnamefont {P.~M.}\ \bibnamefont {Petropoulos}},\ and\ \bibinfo
  {author} {\bibfnamefont {K.}~\bibnamefont {Siampos}},\ }\bibfield  {title}
  {\bibinfo {title} {{Covariant Galilean versus Carrollian hydrodynamics from
  relativistic fluids}},\ }\href {https://doi.org/10.1088/1361-6382/aacf1a}
  {\bibfield  {journal} {\bibinfo  {journal} {Class. Quant. Grav.}\ }\textbf
  {\bibinfo {volume} {35}},\ \bibinfo {pages} {165001} (\bibinfo {year}
  {2018}{\natexlab{b}})},\ \Eprint {https://arxiv.org/abs/1802.05286}
  {arXiv:1802.05286 [hep-th]} \BibitemShut {NoStop}%
\bibitem [{\citenamefont {Campoleoni}\ \emph {et~al.}(2019)\citenamefont
  {Campoleoni}, \citenamefont {Ciambelli}, \citenamefont {Marteau},
  \citenamefont {Petropoulos},\ and\ \citenamefont
  {Siampos}}]{campoleoni2019two}%
  \BibitemOpen
  \bibfield  {author} {\bibinfo {author} {\bibfnamefont {A.}~\bibnamefont
  {Campoleoni}}, \bibinfo {author} {\bibfnamefont {L.}~\bibnamefont
  {Ciambelli}}, \bibinfo {author} {\bibfnamefont {C.}~\bibnamefont {Marteau}},
  \bibinfo {author} {\bibfnamefont {P.~M.}\ \bibnamefont {Petropoulos}},\ and\
  \bibinfo {author} {\bibfnamefont {K.}~\bibnamefont {Siampos}},\ }\bibfield
  {title} {\bibinfo {title} {Two-dimensional fluids and their holographic
  duals},\ }\href@noop {} {\bibfield  {journal} {\bibinfo  {journal} {Nuclear
  Physics B}\ }\textbf {\bibinfo {volume} {946}},\ \bibinfo {pages} {114692}
  (\bibinfo {year} {2019})}\BibitemShut {NoStop}%
\bibitem [{\citenamefont {Ciambelli}\ \emph {et~al.}(2020)\citenamefont
  {Ciambelli}, \citenamefont {Marteau}, \citenamefont {Petropoulos},\ and\
  \citenamefont {Ruzziconi}}]{Ciambelli:2020eba}%
  \BibitemOpen
  \bibfield  {author} {\bibinfo {author} {\bibfnamefont {L.}~\bibnamefont
  {Ciambelli}}, \bibinfo {author} {\bibfnamefont {C.}~\bibnamefont {Marteau}},
  \bibinfo {author} {\bibfnamefont {P.~M.}\ \bibnamefont {Petropoulos}},\ and\
  \bibinfo {author} {\bibfnamefont {R.}~\bibnamefont {Ruzziconi}},\ }\bibfield
  {title} {\bibinfo {title} {{Gauges in Three-Dimensional Gravity and
  Holographic Fluids}},\ }\href {https://doi.org/10.1007/JHEP11(2020)092}
  {\bibfield  {journal} {\bibinfo  {journal} {JHEP}\ }\textbf {\bibinfo
  {volume} {11}},\ \bibinfo {pages} {092}},\ \Eprint
  {https://arxiv.org/abs/2006.10082} {arXiv:2006.10082 [hep-th]} \BibitemShut
  {NoStop}%
\bibitem [{\citenamefont {de~Boer}\ \emph {et~al.}(2020)\citenamefont
  {de~Boer}, \citenamefont {Hartong}, \citenamefont {Have}, \citenamefont
  {Obers},\ and\ \citenamefont {Sybesma}}]{10.21468/SciPostPhys.9.2.018}%
  \BibitemOpen
  \bibfield  {author} {\bibinfo {author} {\bibfnamefont {J.}~\bibnamefont
  {de~Boer}}, \bibinfo {author} {\bibfnamefont {J.}~\bibnamefont {Hartong}},
  \bibinfo {author} {\bibfnamefont {E.}~\bibnamefont {Have}}, \bibinfo {author}
  {\bibfnamefont {N.~A.}\ \bibnamefont {Obers}},\ and\ \bibinfo {author}
  {\bibfnamefont {W.}~\bibnamefont {Sybesma}},\ }\bibfield  {title} {\bibinfo
  {title} {{Non-boost invariant fluid dynamics}},\ }\href
  {https://doi.org/10.21468/SciPostPhys.9.2.018} {\bibfield  {journal}
  {\bibinfo  {journal} {SciPost Phys.}\ }\textbf {\bibinfo {volume} {9}},\
  \bibinfo {pages} {018} (\bibinfo {year} {2020})}\BibitemShut {NoStop}%
\bibitem [{\citenamefont {de~Boer}\ \emph {et~al.}(2022)\citenamefont
  {de~Boer}, \citenamefont {Hartong}, \citenamefont {Obers}, \citenamefont
  {Sybesma},\ and\ \citenamefont {Vandoren}}]{deBoer:2021jej}%
  \BibitemOpen
  \bibfield  {author} {\bibinfo {author} {\bibfnamefont {J.}~\bibnamefont
  {de~Boer}}, \bibinfo {author} {\bibfnamefont {J.}~\bibnamefont {Hartong}},
  \bibinfo {author} {\bibfnamefont {N.~A.}\ \bibnamefont {Obers}}, \bibinfo
  {author} {\bibfnamefont {W.}~\bibnamefont {Sybesma}},\ and\ \bibinfo {author}
  {\bibfnamefont {S.}~\bibnamefont {Vandoren}},\ }\bibfield  {title} {\bibinfo
  {title} {{Carroll Symmetry, Dark Energy and Inflation}},\ }\href
  {https://doi.org/10.3389/fphy.2022.810405} {\bibfield  {journal} {\bibinfo
  {journal} {Front. in Phys.}\ }\textbf {\bibinfo {volume} {10}},\ \bibinfo
  {pages} {810405} (\bibinfo {year} {2022})},\ \Eprint
  {https://arxiv.org/abs/2110.02319} {arXiv:2110.02319 [hep-th]} \BibitemShut
  {NoStop}%
\bibitem [{\citenamefont {Bonga}\ and\ \citenamefont
  {Prabhu}(2020)}]{Bonga:2020fhx}%
  \BibitemOpen
  \bibfield  {author} {\bibinfo {author} {\bibfnamefont {B.}~\bibnamefont
  {Bonga}}\ and\ \bibinfo {author} {\bibfnamefont {K.}~\bibnamefont {Prabhu}},\
  }\bibfield  {title} {\bibinfo {title} {{BMS-like symmetries in cosmology}},\
  }\href {https://doi.org/10.1103/PhysRevD.102.104043} {\bibfield  {journal}
  {\bibinfo  {journal} {Phys. Rev. D}\ }\textbf {\bibinfo {volume} {102}},\
  \bibinfo {pages} {104043} (\bibinfo {year} {2020})},\ \Eprint
  {https://arxiv.org/abs/2009.01243} {arXiv:2009.01243 [gr-qc]} \BibitemShut
  {NoStop}%
\bibitem [{\citenamefont {Bagchi}\ \emph
  {et~al.}(2019{\natexlab{b}})\citenamefont {Bagchi}, \citenamefont
  {Banerjee},\ and\ \citenamefont {Parekh}}]{PhysRevLett.123.111601}%
  \BibitemOpen
  \bibfield  {author} {\bibinfo {author} {\bibfnamefont {A.}~\bibnamefont
  {Bagchi}}, \bibinfo {author} {\bibfnamefont {A.}~\bibnamefont {Banerjee}},\
  and\ \bibinfo {author} {\bibfnamefont {P.}~\bibnamefont {Parekh}},\
  }\bibfield  {title} {\bibinfo {title} {Tensionless path from closed to open
  strings},\ }\href {https://doi.org/10.1103/PhysRevLett.123.111601} {\bibfield
   {journal} {\bibinfo  {journal} {Phys. Rev. Lett.}\ }\textbf {\bibinfo
  {volume} {123}},\ \bibinfo {pages} {111601} (\bibinfo {year}
  {2019}{\natexlab{b}})}\BibitemShut {NoStop}%
\bibitem [{\citenamefont {Bagchi}\ \emph
  {et~al.}(2022{\natexlab{a}})\citenamefont {Bagchi}, \citenamefont {Banerjee},
  \citenamefont {Chakrabortty},\ and\ \citenamefont
  {Chatterjee}}]{Bagchi:2021ban}%
  \BibitemOpen
  \bibfield  {author} {\bibinfo {author} {\bibfnamefont {A.}~\bibnamefont
  {Bagchi}}, \bibinfo {author} {\bibfnamefont {A.}~\bibnamefont {Banerjee}},
  \bibinfo {author} {\bibfnamefont {S.}~\bibnamefont {Chakrabortty}},\ and\
  \bibinfo {author} {\bibfnamefont {R.}~\bibnamefont {Chatterjee}},\ }\bibfield
   {title} {\bibinfo {title} {{A Rindler road to Carrollian worldsheets}},\
  }\href {https://doi.org/10.1007/JHEP04(2022)082} {\bibfield  {journal}
  {\bibinfo  {journal} {JHEP}\ }\textbf {\bibinfo {volume} {04}},\ \bibinfo
  {pages} {082}},\ \Eprint {https://arxiv.org/abs/2111.01172} {arXiv:2111.01172
  [hep-th]} \BibitemShut {NoStop}%
\bibitem [{\citenamefont {Cardona}\ \emph {et~al.}(2016)\citenamefont
  {Cardona}, \citenamefont {Gomis},\ and\ \citenamefont
  {Pons}}]{Cardona:2016ytk}%
  \BibitemOpen
  \bibfield  {author} {\bibinfo {author} {\bibfnamefont {B.}~\bibnamefont
  {Cardona}}, \bibinfo {author} {\bibfnamefont {J.}~\bibnamefont {Gomis}},\
  and\ \bibinfo {author} {\bibfnamefont {J.~M.}\ \bibnamefont {Pons}},\
  }\bibfield  {title} {\bibinfo {title} {{Dynamics of Carroll Strings}},\
  }\href {https://doi.org/10.1007/JHEP07(2016)050} {\bibfield  {journal}
  {\bibinfo  {journal} {JHEP}\ }\textbf {\bibinfo {volume} {07}},\ \bibinfo
  {pages} {050}},\ \Eprint {https://arxiv.org/abs/1605.05483} {arXiv:1605.05483
  [hep-th]} \BibitemShut {NoStop}%
\bibitem [{\citenamefont {P\'erez}(2021)}]{Perez:2021abf}%
  \BibitemOpen
  \bibfield  {author} {\bibinfo {author} {\bibfnamefont {A.}~\bibnamefont
  {P\'erez}},\ }\bibfield  {title} {\bibinfo {title} {{Asymptotic symmetries in
  Carrollian theories of gravity}},\ }\href
  {https://doi.org/10.1007/JHEP12(2021)173} {\bibfield  {journal} {\bibinfo
  {journal} {JHEP}\ }\textbf {\bibinfo {volume} {12}},\ \bibinfo {pages}
  {173}},\ \Eprint {https://arxiv.org/abs/2110.15834} {arXiv:2110.15834
  [hep-th]} \BibitemShut {NoStop}%
\bibitem [{\citenamefont {P\'erez}(2022)}]{Perez:2022jpr}%
  \BibitemOpen
  \bibfield  {author} {\bibinfo {author} {\bibfnamefont {A.}~\bibnamefont
  {P\'erez}},\ }\bibfield  {title} {\bibinfo {title} {{Asymptotic symmetries in
  Carrollian theories of gravity with a negative cosmological constant}},\
  }\href {https://doi.org/10.1007/JHEP09(2022)044} {\bibfield  {journal}
  {\bibinfo  {journal} {JHEP}\ }\textbf {\bibinfo {volume} {09}},\ \bibinfo
  {pages} {044}},\ \Eprint {https://arxiv.org/abs/2202.08768} {arXiv:2202.08768
  [hep-th]} \BibitemShut {NoStop}%
\bibitem [{\citenamefont {Hartong}(2015)}]{Hartong:2015xda}%
  \BibitemOpen
  \bibfield  {author} {\bibinfo {author} {\bibfnamefont {J.}~\bibnamefont
  {Hartong}},\ }\bibfield  {title} {\bibinfo {title} {{Gauging the Carroll
  Algebra and Ultra-Relativistic Gravity}},\ }\href
  {https://doi.org/10.1007/JHEP08(2015)069} {\bibfield  {journal} {\bibinfo
  {journal} {JHEP}\ }\textbf {\bibinfo {volume} {08}},\ \bibinfo {pages}
  {069}},\ \Eprint {https://arxiv.org/abs/1505.05011} {arXiv:1505.05011
  [hep-th]} \BibitemShut {NoStop}%
\bibitem [{\citenamefont {Figueroa-O'Farrill}\ \emph
  {et~al.}(2022{\natexlab{a}})\citenamefont {Figueroa-O'Farrill}, \citenamefont
  {Have}, \citenamefont {Prohazka},\ and\ \citenamefont
  {Salzer}}]{Figueroa-OFarrill:2021sxz}%
  \BibitemOpen
  \bibfield  {author} {\bibinfo {author} {\bibfnamefont {J.}~\bibnamefont
  {Figueroa-O'Farrill}}, \bibinfo {author} {\bibfnamefont {E.}~\bibnamefont
  {Have}}, \bibinfo {author} {\bibfnamefont {S.}~\bibnamefont {Prohazka}},\
  and\ \bibinfo {author} {\bibfnamefont {J.}~\bibnamefont {Salzer}},\
  }\bibfield  {title} {\bibinfo {title} {{Carrollian and celestial spaces at
  infinity}},\ }\href {https://doi.org/10.1007/JHEP09(2022)007} {\bibfield
  {journal} {\bibinfo  {journal} {JHEP}\ }\textbf {\bibinfo {volume} {09}},\
  \bibinfo {pages} {007}},\ \Eprint {https://arxiv.org/abs/2112.03319}
  {arXiv:2112.03319 [hep-th]} \BibitemShut {NoStop}%
\bibitem [{\citenamefont {Hansen}\ \emph {et~al.}(2022)\citenamefont {Hansen},
  \citenamefont {Obers}, \citenamefont {Oling},\ and\ \citenamefont
  {S\o{}gaard}}]{Hansen:2021fxi}%
  \BibitemOpen
  \bibfield  {author} {\bibinfo {author} {\bibfnamefont {D.}~\bibnamefont
  {Hansen}}, \bibinfo {author} {\bibfnamefont {N.~A.}\ \bibnamefont {Obers}},
  \bibinfo {author} {\bibfnamefont {G.}~\bibnamefont {Oling}},\ and\ \bibinfo
  {author} {\bibfnamefont {B.~T.}\ \bibnamefont {S\o{}gaard}},\ }\bibfield
  {title} {\bibinfo {title} {{Carroll Expansion of General Relativity}},\
  }\href {https://doi.org/10.21468/SciPostPhys.13.3.055} {\bibfield  {journal}
  {\bibinfo  {journal} {SciPost Phys.}\ }\textbf {\bibinfo {volume} {13}},\
  \bibinfo {pages} {055} (\bibinfo {year} {2022})},\ \Eprint
  {https://arxiv.org/abs/2112.12684} {arXiv:2112.12684 [hep-th]} \BibitemShut
  {NoStop}%
\bibitem [{\citenamefont {Gomis}\ \emph {et~al.}(2021)\citenamefont {Gomis},
  \citenamefont {Hidalgo},\ and\ \citenamefont
  {Salgado-Rebolledo}}]{Gomis:2020wxp}%
  \BibitemOpen
  \bibfield  {author} {\bibinfo {author} {\bibfnamefont {J.}~\bibnamefont
  {Gomis}}, \bibinfo {author} {\bibfnamefont {D.}~\bibnamefont {Hidalgo}},\
  and\ \bibinfo {author} {\bibfnamefont {P.}~\bibnamefont
  {Salgado-Rebolledo}},\ }\bibfield  {title} {\bibinfo {title}
  {{Non-relativistic and Carrollian limits of Jackiw-Teitelboim gravity}},\
  }\href {https://doi.org/10.1007/JHEP05(2021)162} {\bibfield  {journal}
  {\bibinfo  {journal} {JHEP}\ }\textbf {\bibinfo {volume} {05}},\ \bibinfo
  {pages} {162}},\ \Eprint {https://arxiv.org/abs/2011.15053} {arXiv:2011.15053
  [hep-th]} \BibitemShut {NoStop}%
\bibitem [{\citenamefont {Bergshoeff}\ \emph {et~al.}(2023)\citenamefont
  {Bergshoeff}, \citenamefont {Gomis},\ and\ \citenamefont
  {Kleinschmidt}}]{Bergshoeff:2022qkx}%
  \BibitemOpen
  \bibfield  {author} {\bibinfo {author} {\bibfnamefont {E.~A.}\ \bibnamefont
  {Bergshoeff}}, \bibinfo {author} {\bibfnamefont {J.}~\bibnamefont {Gomis}},\
  and\ \bibinfo {author} {\bibfnamefont {A.}~\bibnamefont {Kleinschmidt}},\
  }\bibfield  {title} {\bibinfo {title} {{Non-Lorentzian theories with and
  without constraints}},\ }\href {https://doi.org/10.1007/JHEP01(2023)167}
  {\bibfield  {journal} {\bibinfo  {journal} {JHEP}\ }\textbf {\bibinfo
  {volume} {01}},\ \bibinfo {pages} {167}},\ \Eprint
  {https://arxiv.org/abs/2210.14848} {arXiv:2210.14848 [hep-th]} \BibitemShut
  {NoStop}%
\bibitem [{\citenamefont {Guerrieri}\ and\ \citenamefont
  {Sobreiro}(2021)}]{Guerrieri:2021cdz}%
  \BibitemOpen
  \bibfield  {author} {\bibinfo {author} {\bibfnamefont {A.}~\bibnamefont
  {Guerrieri}}\ and\ \bibinfo {author} {\bibfnamefont {R.~F.}\ \bibnamefont
  {Sobreiro}},\ }\bibfield  {title} {\bibinfo {title} {{Carroll limit of
  four-dimensional gravity theories in the first order formalism}},\ }\href
  {https://doi.org/10.1088/1361-6382/ac345f} {\bibfield  {journal} {\bibinfo
  {journal} {Class. Quant. Grav.}\ }\textbf {\bibinfo {volume} {38}},\ \bibinfo
  {pages} {245003} (\bibinfo {year} {2021})},\ \Eprint
  {https://arxiv.org/abs/2107.10129} {arXiv:2107.10129 [gr-qc]} \BibitemShut
  {NoStop}%
\bibitem [{\citenamefont {Hansen}\ \emph {et~al.}(2021)\citenamefont {Hansen},
  \citenamefont {Hartong}, \citenamefont {Obers},\ and\ \citenamefont
  {Oling}}]{Hansen:2020wqw}%
  \BibitemOpen
  \bibfield  {author} {\bibinfo {author} {\bibfnamefont {D.}~\bibnamefont
  {Hansen}}, \bibinfo {author} {\bibfnamefont {J.}~\bibnamefont {Hartong}},
  \bibinfo {author} {\bibfnamefont {N.~A.}\ \bibnamefont {Obers}},\ and\
  \bibinfo {author} {\bibfnamefont {G.}~\bibnamefont {Oling}},\ }\bibfield
  {title} {\bibinfo {title} {{Galilean first-order formulation for the
  nonrelativistic expansion of general relativity}},\ }\href
  {https://doi.org/10.1103/PhysRevD.104.L061501} {\bibfield  {journal}
  {\bibinfo  {journal} {Phys. Rev. D}\ }\textbf {\bibinfo {volume} {104}},\
  \bibinfo {pages} {L061501} (\bibinfo {year} {2021})},\ \Eprint
  {https://arxiv.org/abs/2012.01518} {arXiv:2012.01518 [hep-th]} \BibitemShut
  {NoStop}%
\bibitem [{\citenamefont {Anderson}(2004)}]{Anderson:2002zn}%
  \BibitemOpen
  \bibfield  {author} {\bibinfo {author} {\bibfnamefont {E.}~\bibnamefont
  {Anderson}},\ }\bibfield  {title} {\bibinfo {title} {{Strong coupled
  relativity without relativity}},\ }\href
  {https://doi.org/10.1023/B:GERG.0000010474.63835.2c} {\bibfield  {journal}
  {\bibinfo  {journal} {Gen. Rel. Grav.}\ }\textbf {\bibinfo {volume} {36}},\
  \bibinfo {pages} {255} (\bibinfo {year} {2004})},\ \Eprint
  {https://arxiv.org/abs/gr-qc/0205118} {arXiv:gr-qc/0205118} \BibitemShut
  {NoStop}%
\bibitem [{\citenamefont {Donnay}\ and\ \citenamefont
  {Marteau}(2019)}]{Donnay:2019jiz}%
  \BibitemOpen
  \bibfield  {author} {\bibinfo {author} {\bibfnamefont {L.}~\bibnamefont
  {Donnay}}\ and\ \bibinfo {author} {\bibfnamefont {C.}~\bibnamefont
  {Marteau}},\ }\bibfield  {title} {\bibinfo {title} {{Carrollian Physics at
  the Black Hole Horizon}},\ }\href {https://doi.org/10.1088/1361-6382/ab2fd5}
  {\bibfield  {journal} {\bibinfo  {journal} {Class. Quant. Grav.}\ }\textbf
  {\bibinfo {volume} {36}},\ \bibinfo {pages} {165002} (\bibinfo {year}
  {2019})},\ \Eprint {https://arxiv.org/abs/1903.09654} {arXiv:1903.09654
  [hep-th]} \BibitemShut {NoStop}%
\bibitem [{\citenamefont {Grumiller}\ and\ \citenamefont
  {Merbis}(2020)}]{Grumiller:2019tyl}%
  \BibitemOpen
  \bibfield  {author} {\bibinfo {author} {\bibfnamefont {D.}~\bibnamefont
  {Grumiller}}\ and\ \bibinfo {author} {\bibfnamefont {W.}~\bibnamefont
  {Merbis}},\ }\bibfield  {title} {\bibinfo {title} {{Near horizon dynamics of
  three dimensional black holes}},\ }\href
  {https://doi.org/10.21468/SciPostPhys.8.1.010} {\bibfield  {journal}
  {\bibinfo  {journal} {SciPost Phys.}\ }\textbf {\bibinfo {volume} {8}},\
  \bibinfo {pages} {010} (\bibinfo {year} {2020})},\ \Eprint
  {https://arxiv.org/abs/1906.10694} {arXiv:1906.10694 [hep-th]} \BibitemShut
  {NoStop}%
\bibitem [{\citenamefont {Redondo-Yuste}\ and\ \citenamefont
  {Lehner}(2022)}]{Redondo-Yuste:2022czg}%
  \BibitemOpen
  \bibfield  {author} {\bibinfo {author} {\bibfnamefont {J.}~\bibnamefont
  {Redondo-Yuste}}\ and\ \bibinfo {author} {\bibfnamefont {L.}~\bibnamefont
  {Lehner}},\ }\href@noop {} {\bibinfo {title} {{Non-linear black hole dynamics
  and Carrollian fluids}}} (\bibinfo {year} {2022}),\ \Eprint
  {https://arxiv.org/abs/2212.06175} {arXiv:2212.06175 [gr-qc]} \BibitemShut
  {NoStop}%
\bibitem [{\citenamefont {Anabal\'on}\ \emph {et~al.}(2021)\citenamefont
  {Anabal\'on}, \citenamefont {Brenner}, \citenamefont {Giribet},\ and\
  \citenamefont {Montecchio}}]{Anabalon:2021wjy}%
  \BibitemOpen
  \bibfield  {author} {\bibinfo {author} {\bibfnamefont {A.}~\bibnamefont
  {Anabal\'on}}, \bibinfo {author} {\bibfnamefont {S.}~\bibnamefont {Brenner}},
  \bibinfo {author} {\bibfnamefont {G.}~\bibnamefont {Giribet}},\ and\ \bibinfo
  {author} {\bibfnamefont {L.}~\bibnamefont {Montecchio}},\ }\bibfield  {title}
  {\bibinfo {title} {{Closer look at black hole pair creation}},\ }\href
  {https://doi.org/10.1103/PhysRevD.104.024044} {\bibfield  {journal} {\bibinfo
   {journal} {Phys. Rev. D}\ }\textbf {\bibinfo {volume} {104}},\ \bibinfo
  {pages} {024044} (\bibinfo {year} {2021})},\ \Eprint
  {https://arxiv.org/abs/2103.05782} {arXiv:2103.05782 [hep-th]} \BibitemShut
  {NoStop}%
\bibitem [{\citenamefont {Ecker}\ \emph {et~al.}(2023)\citenamefont {Ecker},
  \citenamefont {Grumiller}, \citenamefont {Hartong}, \citenamefont {P\'erez},
  \citenamefont {Prohazka},\ and\ \citenamefont {Troncoso}}]{Ecker:2023uwm}%
  \BibitemOpen
  \bibfield  {author} {\bibinfo {author} {\bibfnamefont {F.}~\bibnamefont
  {Ecker}}, \bibinfo {author} {\bibfnamefont {D.}~\bibnamefont {Grumiller}},
  \bibinfo {author} {\bibfnamefont {J.}~\bibnamefont {Hartong}}, \bibinfo
  {author} {\bibfnamefont {A.}~\bibnamefont {P\'erez}}, \bibinfo {author}
  {\bibfnamefont {S.}~\bibnamefont {Prohazka}},\ and\ \bibinfo {author}
  {\bibfnamefont {R.}~\bibnamefont {Troncoso}},\ }\bibfield  {title} {\bibinfo
  {title} {{Carroll black holes}},\ }\href@noop {} {\  (\bibinfo {year}
  {2023})},\ \Eprint {https://arxiv.org/abs/2308.10947} {arXiv:2308.10947
  [hep-th]} \BibitemShut {NoStop}%
\bibitem [{\citenamefont {Herfray}(2022)}]{Herfray:2021qmp}%
  \BibitemOpen
  \bibfield  {author} {\bibinfo {author} {\bibfnamefont {Y.}~\bibnamefont
  {Herfray}},\ }\bibfield  {title} {\bibinfo {title} {{Carrollian manifolds and
  null infinity: a view from Cartan geometry}},\ }\href
  {https://doi.org/10.1088/1361-6382/ac635f} {\bibfield  {journal} {\bibinfo
  {journal} {Class. Quant. Grav.}\ }\textbf {\bibinfo {volume} {39}},\ \bibinfo
  {pages} {215005} (\bibinfo {year} {2022})},\ \Eprint
  {https://arxiv.org/abs/2112.09048} {arXiv:2112.09048 [gr-qc]} \BibitemShut
  {NoStop}%
\bibitem [{\citenamefont {Chandrasekaran}\ \emph {et~al.}(2022)\citenamefont
  {Chandrasekaran}, \citenamefont {Flanagan}, \citenamefont {Shehzad},\ and\
  \citenamefont {Speranza}}]{Chandrasekaran:2021hxc}%
  \BibitemOpen
  \bibfield  {author} {\bibinfo {author} {\bibfnamefont {V.}~\bibnamefont
  {Chandrasekaran}}, \bibinfo {author} {\bibfnamefont {E.~E.}\ \bibnamefont
  {Flanagan}}, \bibinfo {author} {\bibfnamefont {I.}~\bibnamefont {Shehzad}},\
  and\ \bibinfo {author} {\bibfnamefont {A.~J.}\ \bibnamefont {Speranza}},\
  }\bibfield  {title} {\bibinfo {title} {{Brown-York charges at null
  boundaries}},\ }\href {https://doi.org/10.1007/JHEP01(2022)029} {\bibfield
  {journal} {\bibinfo  {journal} {JHEP}\ }\textbf {\bibinfo {volume} {01}},\
  \bibinfo {pages} {029}},\ \Eprint {https://arxiv.org/abs/2109.11567}
  {arXiv:2109.11567 [hep-th]} \BibitemShut {NoStop}%
\bibitem [{\citenamefont {Ciambelli}\ \emph {et~al.}(2019)\citenamefont
  {Ciambelli}, \citenamefont {Leigh}, \citenamefont {Marteau},\ and\
  \citenamefont {Petropoulos}}]{Ciambelli:2019lap}%
  \BibitemOpen
  \bibfield  {author} {\bibinfo {author} {\bibfnamefont {L.}~\bibnamefont
  {Ciambelli}}, \bibinfo {author} {\bibfnamefont {R.~G.}\ \bibnamefont
  {Leigh}}, \bibinfo {author} {\bibfnamefont {C.}~\bibnamefont {Marteau}},\
  and\ \bibinfo {author} {\bibfnamefont {P.~M.}\ \bibnamefont {Petropoulos}},\
  }\bibfield  {title} {\bibinfo {title} {{Carroll Structures, Null Geometry and
  Conformal Isometries}},\ }\href {https://doi.org/10.1103/PhysRevD.100.046010}
  {\bibfield  {journal} {\bibinfo  {journal} {Phys. Rev. D}\ }\textbf {\bibinfo
  {volume} {100}},\ \bibinfo {pages} {046010} (\bibinfo {year} {2019})},\
  \Eprint {https://arxiv.org/abs/1905.02221} {arXiv:1905.02221 [hep-th]}
  \BibitemShut {NoStop}%
\bibitem [{\citenamefont {Gray}\ \emph {et~al.}(2022)\citenamefont {Gray},
  \citenamefont {Kubiznak}, \citenamefont {Perche},\ and\ \citenamefont
  {Redondo-Yuste}}]{Gray:2022svz}%
  \BibitemOpen
  \bibfield  {author} {\bibinfo {author} {\bibfnamefont {F.}~\bibnamefont
  {Gray}}, \bibinfo {author} {\bibfnamefont {D.}~\bibnamefont {Kubiznak}},
  \bibinfo {author} {\bibfnamefont {T.~R.}\ \bibnamefont {Perche}},\ and\
  \bibinfo {author} {\bibfnamefont {J.}~\bibnamefont {Redondo-Yuste}},\
  }\href@noop {} {\bibinfo {title} {{Carrollian Motion in Magnetized Black Hole
  Horizons}}} (\bibinfo {year} {2022}),\ \Eprint
  {https://arxiv.org/abs/2211.13695} {arXiv:2211.13695 [gr-qc]} \BibitemShut
  {NoStop}%
\bibitem [{\citenamefont {Bicak}\ \emph {et~al.}(2023)\citenamefont {Bicak},
  \citenamefont {Kubiznak},\ and\ \citenamefont {Perche}}]{Bicak:2023vxs}%
  \BibitemOpen
  \bibfield  {author} {\bibinfo {author} {\bibfnamefont {J.}~\bibnamefont
  {Bicak}}, \bibinfo {author} {\bibfnamefont {D.}~\bibnamefont {Kubiznak}},\
  and\ \bibinfo {author} {\bibfnamefont {T.~R.}\ \bibnamefont {Perche}},\
  }\href@noop {} {\bibinfo {title} {{Monarch Migration of Carrollian Particles
  on the Black Hole Horizon}}} (\bibinfo {year} {2023}),\ \Eprint
  {https://arxiv.org/abs/2302.11639} {arXiv:2302.11639 [gr-qc]} \BibitemShut
  {NoStop}%
\bibitem [{\citenamefont {Marsot}\ \emph
  {et~al.}(2022{\natexlab{b}})\citenamefont {Marsot}, \citenamefont {Zhang},\
  and\ \citenamefont {Horvathy}}]{Marsot:2022qkx}%
  \BibitemOpen
  \bibfield  {author} {\bibinfo {author} {\bibfnamefont {L.}~\bibnamefont
  {Marsot}}, \bibinfo {author} {\bibfnamefont {P.-M.}\ \bibnamefont {Zhang}},\
  and\ \bibinfo {author} {\bibfnamefont {P.}~\bibnamefont {Horvathy}},\
  }\bibfield  {title} {\bibinfo {title} {{Anyonic spin-Hall effect on the black
  hole horizon}},\ }\href {https://doi.org/10.1103/PhysRevD.106.L121503}
  {\bibfield  {journal} {\bibinfo  {journal} {Phys. Rev. D}\ }\textbf {\bibinfo
  {volume} {106}},\ \bibinfo {pages} {L121503} (\bibinfo {year}
  {2022}{\natexlab{b}})},\ \Eprint {https://arxiv.org/abs/2207.06302}
  {arXiv:2207.06302 [gr-qc]} \BibitemShut {NoStop}%
\bibitem [{\citenamefont {Price}\ and\ \citenamefont
  {Thorne}(1986)}]{Price:1986yy}%
  \BibitemOpen
  \bibfield  {author} {\bibinfo {author} {\bibfnamefont {R.~H.}\ \bibnamefont
  {Price}}\ and\ \bibinfo {author} {\bibfnamefont {K.~S.}\ \bibnamefont
  {Thorne}},\ }\bibfield  {title} {\bibinfo {title} {{Membrane Viewpoint on
  Black Holes: Properties and Evolution of the Stretched Horizon}},\ }\href
  {https://doi.org/10.1103/PhysRevD.33.915} {\bibfield  {journal} {\bibinfo
  {journal} {Phys. Rev. D}\ }\textbf {\bibinfo {volume} {33}},\ \bibinfo
  {pages} {915} (\bibinfo {year} {1986})}\BibitemShut {NoStop}%
\bibitem [{\citenamefont {{Thorne}}\ \emph {et~al.}(1986)\citenamefont
  {{Thorne}}, \citenamefont {{Price}},\ and\ \citenamefont
  {{MacDonald}}}]{1986bhmp.book.....T}%
  \BibitemOpen
  \bibfield  {author} {\bibinfo {author} {\bibfnamefont {K.~S.}\ \bibnamefont
  {{Thorne}}}, \bibinfo {author} {\bibfnamefont {R.~H.}\ \bibnamefont
  {{Price}}},\ and\ \bibinfo {author} {\bibfnamefont {D.~A.}\ \bibnamefont
  {{MacDonald}}},\ }\href@noop {} {\emph {\bibinfo {title} {{Black holes: The
  membrane paradigm}}}}\ (\bibinfo {year} {1986})\BibitemShut {NoStop}%
\bibitem [{\citenamefont {Damour}(1978)}]{Damour:1978cg}%
  \BibitemOpen
  \bibfield  {author} {\bibinfo {author} {\bibfnamefont {T.}~\bibnamefont
  {Damour}},\ }\bibfield  {title} {\bibinfo {title} {{Black Hole Eddy
  Currents}},\ }\href {https://doi.org/10.1103/PhysRevD.18.3598} {\bibfield
  {journal} {\bibinfo  {journal} {Phys. Rev. D}\ }\textbf {\bibinfo {volume}
  {18}},\ \bibinfo {pages} {3598} (\bibinfo {year} {1978})}\BibitemShut
  {NoStop}%
\bibitem [{\citenamefont {Henneaux}(1979)}]{Henneaux:1979vn}%
  \BibitemOpen
  \bibfield  {author} {\bibinfo {author} {\bibfnamefont {M.}~\bibnamefont
  {Henneaux}},\ }\bibfield  {title} {\bibinfo {title} {{Geometry of Zero
  Signature Space-times}},\ }\href@noop {} {\bibfield  {journal} {\bibinfo
  {journal} {Bull. Soc. Math. Belg.}\ }\textbf {\bibinfo {volume} {31}},\
  \bibinfo {pages} {47} (\bibinfo {year} {1979})}\BibitemShut {NoStop}%
\bibitem [{\citenamefont {Campoleoni}\ \emph {et~al.}(2022)\citenamefont
  {Campoleoni}, \citenamefont {Henneaux}, \citenamefont {Pekar}, \citenamefont
  {P\'erez},\ and\ \citenamefont {Salgado-Rebolledo}}]{Campoleoni:2022ebj}%
  \BibitemOpen
  \bibfield  {author} {\bibinfo {author} {\bibfnamefont {A.}~\bibnamefont
  {Campoleoni}}, \bibinfo {author} {\bibfnamefont {M.}~\bibnamefont
  {Henneaux}}, \bibinfo {author} {\bibfnamefont {S.}~\bibnamefont {Pekar}},
  \bibinfo {author} {\bibfnamefont {A.}~\bibnamefont {P\'erez}},\ and\ \bibinfo
  {author} {\bibfnamefont {P.}~\bibnamefont {Salgado-Rebolledo}},\ }\bibfield
  {title} {\bibinfo {title} {{Magnetic Carrollian gravity from the Carroll
  algebra}},\ }\href {https://doi.org/10.1007/JHEP09(2022)127} {\bibfield
  {journal} {\bibinfo  {journal} {JHEP}\ }\textbf {\bibinfo {volume} {09}},\
  \bibinfo {pages} {127}},\ \Eprint {https://arxiv.org/abs/2207.14167}
  {arXiv:2207.14167 [hep-th]} \BibitemShut {NoStop}%
\bibitem [{\citenamefont {Niedermaier}(2023)}]{Niedermaier:2023hgk}%
  \BibitemOpen
  \bibfield  {author} {\bibinfo {author} {\bibfnamefont {M.}~\bibnamefont
  {Niedermaier}},\ }\bibfield  {title} {\bibinfo {title} {{Higher derivative
  gravity\textquoteright{}s anti-Newtonian limit and the
  Caldirola\textendash{}Kanai oscillator}},\ }\href
  {https://doi.org/10.1088/1361-6382/acaae4} {\bibfield  {journal} {\bibinfo
  {journal} {Class. Quant. Grav.}\ }\textbf {\bibinfo {volume} {40}},\ \bibinfo
  {pages} {025017} (\bibinfo {year} {2023})}\BibitemShut {NoStop}%
\bibitem [{\citenamefont {Abebe}(2014)}]{Abebe:2014hka}%
  \BibitemOpen
  \bibfield  {author} {\bibinfo {author} {\bibfnamefont {A.}~\bibnamefont
  {Abebe}},\ }\bibfield  {title} {\bibinfo {title} {{Anti-Newtonian cosmologies
  in f(R) gravity}},\ }\href {https://doi.org/10.1088/0264-9381/31/11/115011}
  {\bibfield  {journal} {\bibinfo  {journal} {Class. Quant. Grav.}\ }\textbf
  {\bibinfo {volume} {31}},\ \bibinfo {pages} {115011} (\bibinfo {year}
  {2014})},\ \Eprint {https://arxiv.org/abs/1401.3596} {arXiv:1401.3596
  [gr-qc]} \BibitemShut {NoStop}%
\bibitem [{\citenamefont {Abebe}(2016)}]{Abebe:2016obi}%
  \BibitemOpen
  \bibfield  {author} {\bibinfo {author} {\bibfnamefont {A.}~\bibnamefont
  {Abebe}},\ }\bibfield  {title} {\bibinfo {title} {{Existence of
  anti-Newtonian solutions in fourth-order gravity}},\ }in\ \href@noop {}
  {\emph {\bibinfo {booktitle} {{61st Annual Conference of the South African
  Institute of Physics}}}}\ (\bibinfo {year} {2016})\ pp.\ \bibinfo {pages}
  {201--206}\BibitemShut {NoStop}%
\bibitem [{\citenamefont {Abebe}\ and\ \citenamefont
  {Elmardi}(2015)}]{abebe2015irrotational}%
  \BibitemOpen
  \bibfield  {author} {\bibinfo {author} {\bibfnamefont {A.}~\bibnamefont
  {Abebe}}\ and\ \bibinfo {author} {\bibfnamefont {M.}~\bibnamefont
  {Elmardi}},\ }\bibfield  {title} {\bibinfo {title} {Irrotational-fluid
  cosmologies in fourth-order gravity},\ }\href@noop {} {\bibfield  {journal}
  {\bibinfo  {journal} {International Journal of Geometric Methods in Modern
  Physics}\ }\textbf {\bibinfo {volume} {12}},\ \bibinfo {pages} {1550118}
  (\bibinfo {year} {2015})}\BibitemShut {NoStop}%
\bibitem [{\citenamefont {de~la Cruz-Dombriz}\ and\ \citenamefont
  {Dobado}(2006)}]{delaCruz-Dombriz:2006kob}%
  \BibitemOpen
  \bibfield  {author} {\bibinfo {author} {\bibfnamefont {A.}~\bibnamefont
  {de~la Cruz-Dombriz}}\ and\ \bibinfo {author} {\bibfnamefont
  {A.}~\bibnamefont {Dobado}},\ }\bibfield  {title} {\bibinfo {title} {{A f(R)
  gravity without cosmological constant}},\ }\href
  {https://doi.org/10.1103/PhysRevD.74.087501} {\bibfield  {journal} {\bibinfo
  {journal} {Phys. Rev. D}\ }\textbf {\bibinfo {volume} {74}},\ \bibinfo
  {pages} {087501} (\bibinfo {year} {2006})},\ \Eprint
  {https://arxiv.org/abs/gr-qc/0607118} {arXiv:gr-qc/0607118} \BibitemShut
  {NoStop}%
\bibitem [{\citenamefont {Inonu}\ and\ \citenamefont
  {Wigner}(1953)}]{4860f44e-649d-341b-9a70-b912b6531bea}%
  \BibitemOpen
  \bibfield  {author} {\bibinfo {author} {\bibfnamefont {E.}~\bibnamefont
  {Inonu}}\ and\ \bibinfo {author} {\bibfnamefont {E.~P.}\ \bibnamefont
  {Wigner}},\ }\bibfield  {title} {\bibinfo {title} {On the contraction of
  groups and their representations},\ }\href
  {http://www.jstor.org/stable/88703} {\bibfield  {journal} {\bibinfo
  {journal} {Proceedings of the National Academy of Sciences of the United
  States of America}\ }\textbf {\bibinfo {volume} {39}},\ \bibinfo {pages}
  {510} (\bibinfo {year} {1953})}\BibitemShut {NoStop}%
\bibitem [{\citenamefont {Figueroa-O'Farrill}\ \emph
  {et~al.}(2022{\natexlab{b}})\citenamefont {Figueroa-O'Farrill}, \citenamefont
  {Have}, \citenamefont {Prohazka},\ and\ \citenamefont
  {Salzer}}]{Figueroa-OFarrill:2022mcy}%
  \BibitemOpen
  \bibfield  {author} {\bibinfo {author} {\bibfnamefont {J.}~\bibnamefont
  {Figueroa-O'Farrill}}, \bibinfo {author} {\bibfnamefont {E.}~\bibnamefont
  {Have}}, \bibinfo {author} {\bibfnamefont {S.}~\bibnamefont {Prohazka}},\
  and\ \bibinfo {author} {\bibfnamefont {J.}~\bibnamefont {Salzer}},\
  }\bibfield  {title} {\bibinfo {title} {{The gauging procedure and carrollian
  gravity}},\ }\href {https://doi.org/10.1007/JHEP09(2022)243} {\bibfield
  {journal} {\bibinfo  {journal} {JHEP}\ }\textbf {\bibinfo {volume} {09}},\
  \bibinfo {pages} {243}},\ \Eprint {https://arxiv.org/abs/2206.14178}
  {arXiv:2206.14178 [hep-th]} \BibitemShut {NoStop}%
\bibitem [{\citenamefont
  {Figueroa-O'Farrill}(2023)}]{Figueroa-OFarrill:2022pus}%
  \BibitemOpen
  \bibfield  {author} {\bibinfo {author} {\bibfnamefont {J.}~\bibnamefont
  {Figueroa-O'Farrill}},\ }\bibfield  {title} {\bibinfo {title} {{Lie algebraic
  Carroll/Galilei duality}},\ }\href {https://doi.org/10.1063/5.0132661}
  {\bibfield  {journal} {\bibinfo  {journal} {J. Math. Phys.}\ }\textbf
  {\bibinfo {volume} {64}},\ \bibinfo {pages} {013503} (\bibinfo {year}
  {2023})},\ \Eprint {https://arxiv.org/abs/2210.13924} {arXiv:2210.13924
  [math.DG]} \BibitemShut {NoStop}%
\bibitem [{\citenamefont {Freidel}\ and\ \citenamefont
  {Jai-akson}(2022)}]{Freidel:2022vjq}%
  \BibitemOpen
  \bibfield  {author} {\bibinfo {author} {\bibfnamefont {L.}~\bibnamefont
  {Freidel}}\ and\ \bibinfo {author} {\bibfnamefont {P.}~\bibnamefont
  {Jai-akson}},\ }\href@noop {} {\bibinfo {title} {{Carrollian hydrodynamics
  and symplectic structure on stretched horizons}}} (\bibinfo {year} {2022}),\
  \Eprint {https://arxiv.org/abs/2211.06415} {arXiv:2211.06415 [gr-qc]}
  \BibitemShut {NoStop}%
\bibitem [{\citenamefont {Morand}(2020)}]{Morand:2018tke}%
  \BibitemOpen
  \bibfield  {author} {\bibinfo {author} {\bibfnamefont {K.}~\bibnamefont
  {Morand}},\ }\bibfield  {title} {\bibinfo {title} {{Embedding Galilean and
  Carrollian geometries I. Gravitational waves}},\ }\href
  {https://doi.org/10.1063/1.5130907} {\bibfield  {journal} {\bibinfo
  {journal} {J. Math. Phys.}\ }\textbf {\bibinfo {volume} {61}},\ \bibinfo
  {pages} {082502} (\bibinfo {year} {2020})},\ \Eprint
  {https://arxiv.org/abs/1811.12681} {arXiv:1811.12681 [hep-th]} \BibitemShut
  {NoStop}%
\bibitem [{\citenamefont {Bergshoeff}\ \emph {et~al.}(2017)\citenamefont
  {Bergshoeff}, \citenamefont {Gomis}, \citenamefont {Rollier}, \citenamefont
  {Rosseel},\ and\ \citenamefont {ter Veldhuis}}]{Bergshoeff:2017btm}%
  \BibitemOpen
  \bibfield  {author} {\bibinfo {author} {\bibfnamefont {E.}~\bibnamefont
  {Bergshoeff}}, \bibinfo {author} {\bibfnamefont {J.}~\bibnamefont {Gomis}},
  \bibinfo {author} {\bibfnamefont {B.}~\bibnamefont {Rollier}}, \bibinfo
  {author} {\bibfnamefont {J.}~\bibnamefont {Rosseel}},\ and\ \bibinfo {author}
  {\bibfnamefont {T.}~\bibnamefont {ter Veldhuis}},\ }\bibfield  {title}
  {\bibinfo {title} {{Carroll versus Galilei Gravity}},\ }\href
  {https://doi.org/10.1007/JHEP03(2017)165} {\bibfield  {journal} {\bibinfo
  {journal} {JHEP}\ }\textbf {\bibinfo {volume} {03}},\ \bibinfo {pages}
  {165}},\ \Eprint {https://arxiv.org/abs/1701.06156} {arXiv:1701.06156
  [hep-th]} \BibitemShut {NoStop}%
\bibitem [{\citenamefont {Gupta}\ and\ \citenamefont
  {Suryanarayana}(2021)}]{Gupta:2020dtl}%
  \BibitemOpen
  \bibfield  {author} {\bibinfo {author} {\bibfnamefont {N.}~\bibnamefont
  {Gupta}}\ and\ \bibinfo {author} {\bibfnamefont {N.~V.}\ \bibnamefont
  {Suryanarayana}},\ }\bibfield  {title} {\bibinfo {title} {{Constructing
  Carrollian CFTs}},\ }\href {https://doi.org/10.1007/JHEP03(2021)194}
  {\bibfield  {journal} {\bibinfo  {journal} {JHEP}\ }\textbf {\bibinfo
  {volume} {03}},\ \bibinfo {pages} {194}},\ \Eprint
  {https://arxiv.org/abs/2001.03056} {arXiv:2001.03056 [hep-th]} \BibitemShut
  {NoStop}%
\bibitem [{\citenamefont {Bagchi}\ \emph
  {et~al.}(2022{\natexlab{b}})\citenamefont {Bagchi}, \citenamefont
  {Banerjee},\ and\ \citenamefont {Muraki}}]{Bagchi:2022nvj}%
  \BibitemOpen
  \bibfield  {author} {\bibinfo {author} {\bibfnamefont {A.}~\bibnamefont
  {Bagchi}}, \bibinfo {author} {\bibfnamefont {A.}~\bibnamefont {Banerjee}},\
  and\ \bibinfo {author} {\bibfnamefont {H.}~\bibnamefont {Muraki}},\
  }\bibfield  {title} {\bibinfo {title} {{Boosting to BMS}},\ }\href
  {https://doi.org/10.1007/JHEP09(2022)251} {\bibfield  {journal} {\bibinfo
  {journal} {JHEP}\ }\textbf {\bibinfo {volume} {09}},\ \bibinfo {pages}
  {251}},\ \Eprint {https://arxiv.org/abs/2205.05094} {arXiv:2205.05094
  [hep-th]} \BibitemShut {NoStop}%
\bibitem [{\citenamefont {Baiguera}\ \emph {et~al.}(2023)\citenamefont
  {Baiguera}, \citenamefont {Oling}, \citenamefont {Sybesma},\ and\
  \citenamefont {S\o{}gaard}}]{Baiguera:2022lsw}%
  \BibitemOpen
  \bibfield  {author} {\bibinfo {author} {\bibfnamefont {S.}~\bibnamefont
  {Baiguera}}, \bibinfo {author} {\bibfnamefont {G.}~\bibnamefont {Oling}},
  \bibinfo {author} {\bibfnamefont {W.}~\bibnamefont {Sybesma}},\ and\ \bibinfo
  {author} {\bibfnamefont {B.~T.}\ \bibnamefont {S\o{}gaard}},\ }\bibfield
  {title} {\bibinfo {title} {{Conformal Carroll scalars with boosts}},\ }\href
  {https://doi.org/10.21468/SciPostPhys.14.4.086} {\bibfield  {journal}
  {\bibinfo  {journal} {SciPost Phys.}\ }\textbf {\bibinfo {volume} {14}},\
  \bibinfo {pages} {086} (\bibinfo {year} {2023})},\ \Eprint
  {https://arxiv.org/abs/2207.03468} {arXiv:2207.03468 [hep-th]} \BibitemShut
  {NoStop}%
\bibitem [{\citenamefont {Petkou}\ \emph {et~al.}(2022)\citenamefont {Petkou},
  \citenamefont {Petropoulos}, \citenamefont {Betancour},\ and\ \citenamefont
  {Siampos}}]{Petkou:2022bmz}%
  \BibitemOpen
  \bibfield  {author} {\bibinfo {author} {\bibfnamefont {A.~C.}\ \bibnamefont
  {Petkou}}, \bibinfo {author} {\bibfnamefont {P.~M.}\ \bibnamefont
  {Petropoulos}}, \bibinfo {author} {\bibfnamefont {D.~R.}\ \bibnamefont
  {Betancour}},\ and\ \bibinfo {author} {\bibfnamefont {K.}~\bibnamefont
  {Siampos}},\ }\bibfield  {title} {\bibinfo {title} {{Relativistic fluids,
  hydrodynamic frames and their Galilean versus Carrollian avatars}},\ }\href
  {https://doi.org/10.1007/JHEP09(2022)162} {\bibfield  {journal} {\bibinfo
  {journal} {JHEP}\ }\textbf {\bibinfo {volume} {09}},\ \bibinfo {pages}
  {162}},\ \Eprint {https://arxiv.org/abs/2205.09142} {arXiv:2205.09142
  [hep-th]} \BibitemShut {NoStop}%
\end{thebibliography}
\end{document}